\documentclass[12pt]{article}
\usepackage{graphicx}
\usepackage{amsmath,amssymb}
\newcommand{\dbox}{\,\raise2pt\hbox{\fbox{\rule{2.5pt}{0pt}\rule{0pt}{2.5pt}}}\,}
\newcommand{\qed}{\,\raise0pt\hbox{\mbox{\rule{6.5pt}{6.5pt}}}}
\newcommand{\bra}[1]{\mbox{$\langle #1 |$}}
\newcommand{\ket}[1]{\mbox{$| #1 \rangle$}}

\begin{document}
\setlength{\baselineskip}{6.5mm}

\begin{titlepage}
 \begin{normalsize}
  \begin{flushright}
        UT-Komaba/16-3\\
        March 2016
 \end{flushright}
 \end{normalsize}
 \begin{LARGE}
   \vspace{1cm}
   \begin{center}
   Supersymmetric extended string field theory
\\
in ${\rm NS}^{ n}$ sector and ${\rm NS}^{n-1}$-R sector
\\
   \end{center}
 \end{LARGE}
  \vspace{5mm}
 \begin{center}
    Masako {\sc Asano} 
            \hspace{3mm}and\hspace{3mm}
    Mitsuhiro {\sc Kato}$^{\dagger}$ 
\\
      \vspace{4mm}
        {\sl Faculty of Science and Technology}\\
        {\sl Seikei University}\\
        {\sl Musashino-shi, Tokyo 180-8633, Japan}\\
      \vspace{4mm}
        ${}^{\dagger}${\sl Institute of Physics} \\
        {\sl University of Tokyo, Komaba}\\
        {\sl Meguro-ku, Tokyo 153-8902, Japan}\\
      \vspace{1cm}

  ABSTRACT\par
 \end{center}
 \begin{quote}
  \begin{normalsize}
We construct a class of quadratic gauge invariant actions for extended string fields defined on 
the tensor product of open superstring state space for multiple open string Neveu-Schwarz (NS) sectors with or without one Ramond (R) sector.
The basic idea is the same as for the bosonic extended string field theory developed by the authors~\cite{Asano:2013rka}.
The theory for ${\rm NS}^n$ sector and ${\rm NS}^{n-1}$-R sector contains general $n$-th rank tensor fields and 
$(n\!-\!1)$-th rank spinor-tensor fields in the massless spectrum respectively. 
In principle, 
consistent gauge invariant actions for any generic type of 10-dimensional 
massive or massless tensor or spinor-tensor fields can be extracted from the theory.
We discuss some simple examples of bosonic and fermionic massless actions.

\end{normalsize}
 \end{quote}

\end{titlepage}
\vfil\eject

\section{Introduction}
In the previous paper~\cite{Asano:2013rka}, the authors constructed an extended string field theory (ESFT) which describes massless higher spin fields accompanied with a tower of massive fields, in hoping that it may give a possible ultraviolet completion of the higher spin gauge theory. There the key ingredient is a tensor product of open string state space which naturally gives higher spin fields at a massless level provided with the proper restriction of the states as an extension of the $L_0-\bar L_0=0$ condition for the closed string. Although the interaction is still to be studied, both the gauge-invariant and the gauge-fixed free actions of the various types of higher rank tensor fields are systematically extracted from the quadratic action of ESFT.

In the present paper, we extend the above mentioned construction to include fermionic fields i.e.~higher rank spinor-tensor fields as well as bosonic pure tensor fields, especially in a supersymmetric way. Thus we use the NSR formalism with GSO projection to setup building blocks of open string state space. For the sake of brevity we only consider the simplest supersymmetric case where the tensor product space consists of only two sectors, namely NS$\otimes\cdots\otimes$NS and NS$\otimes\cdots\otimes$NS$\otimes$R sectors, so that the resultant theory has ${\cal N}=1$ supersymmetry.

This paper is organized as follows.
In the next section, we briefly review the free covariant open superstring field theory for NS and R sectors. We then show that the `$a$-gauges,' which is a class of covariant gauge fixing conditions valid for bosonic string field theory \cite{Asano:2006hk, Asano:2012qn}, can be extended to the superstring field theory in both sectors.
In the main section~3, we construct free extended string field theory for ${\rm NS}^{n}$  and ${\rm NS}^{n-1}$-R sectors and discuss the properties of the actions. Massless spectrum of these sectors generally includes higher-spin fields since it is given by the $n$-th rank tensor field and the $(n-1)$-th rank spinor-tensor field respectively.
We see that the basic structure of the actions for ${\rm NS}^{n}$ and ${\rm NS}^{n-1}$-R sectors does not depend on $n$. 
We then explicitly see the massless part of the actions and give some examples of gauge invariant actions for several types of tensor or spinor-tensor fields.
We close the section by giving some comments.
In the final section~4, we give summary and some discussions.
In Appendix~A, we summarize the basic properties of open superstring states and operators.

\section{Quadratic action of  superstring field theory in NS and R sectors}
In this section, we first recall the quadratic action of covariant open superstring field theory in NS and R sectors.
We then see how the $a$-gauges~\cite{Asano:2006hk, Asano:2012qn} can be extended to the gauge invariant action in NS and R sectors. 

\subsection{State space and the gauge invariant action}
The state spaces for NS and R sectors we use have the form 
\begin{equation}
{\cal H}_{\rm (NS)} =\tilde{\cal F}^{\rm (NS)} + c_0 \tilde{\cal F}^{\rm (NS)}
,\qquad
 {\cal H}_{\rm (R)} =\tilde{\cal F}^{\rm (R)} + (\gamma_0 + c_0\tilde{G}_0) \tilde{\cal F}^{\rm (R)}.
\label{eq:Rstate}
\end{equation}
Here the spaces $\tilde{\cal F}^{\rm (NS)}$ and $\tilde{\cal F}^{\rm (R)}$ consist of states with arbitrary number of non-zero modes of matter ($\alpha$, $\psi$) and ghost ($b$, $c$, $\beta$, $\gamma$) oscillators operated respectively on the ghost number 1 ground states 
\begin{equation}
\ket{0,p;\downarrow; -1}_{\rm NS} \,(\,=\ket{0,p}\otimes\ket{\downarrow}\otimes\ket{-1} \,),
\qquad
\ket{0,p,a;\downarrow; -\frac{1}{2}}_{\rm R } \, (\, =\ket{0,p,a} \otimes \ket{\downarrow} \otimes\ket{-\frac{1}{2}} \,).
\label{eq:groundst}
\end{equation}
For the R sector, we use the constrained space $ {\cal H}_{\rm (R)} $ following the formulation developed in refs.\cite{Kazama:1986cy, Banks:1985xa, Terao:1986ex, LeClair:1986pi, Date:1986tg}. 
Note that we choose the picture number $-1$ for the NS sector and  $-\frac{1}{2}$ for the R sector as in ref.~\cite{Witten:1986qs}. 
That is, the ground states $\ket{-1}  $ and $\ket{ -\frac{1}{2}} $ satisfy
\begin{equation}
\beta_r \ket{-1} = \gamma_r \ket{-1}=0
\qquad ({\mbox{for $r>0$}}),
\end{equation}
\begin{equation}
\beta_n \ket{ -\frac{1}{2}} =0
\quad (\mbox{for $n \ge 0$}),
\qquad
\gamma_n \ket{ -\frac{1}{2}} =0
\quad ({\mbox{for $n >0$} }).
\end{equation}
The number of independent matter ground states $\ket{0,p,a} $ ($a=1,\cdots, 32$) for the R sector is $2^{D/2}=32$ which is equivalent to the dimension of 10-dimensional spinor. 
We regard the operation of the fermion zero mode $ \tilde{\psi}_0{}^{\mu} \,(= \sqrt{2} {\psi}_0{}^{\mu} )$ on $\ket{0,p,a} $ as
\begin{equation}
\tilde{\psi}_0{}^{\mu} \ket{0,p,a}  =  \ket{0,p,b} \Gamma_{ba}^\mu
\end{equation}
where $\Gamma_{ba}^\mu $ denotes the 10-dimensional gamma matrix.
In particular, for the operator $\tilde{\gamma}^{11}$ defined by the product of all the $\tilde{\psi}_0{}^{\mu}$'s as $\tilde{\gamma}^{11} \equiv \tilde{\psi}_0{}^9 \cdots \tilde{\psi}_0{}^0 $, 
the relation 
\begin{equation}
\tilde{\gamma}^{11} \ket{0,p,a}  =  \ket{0,p,b} \Gamma_{ba}^{11}
\end{equation}
holds for $\Gamma^{11}= \Gamma^{0}\cdots \Gamma^{9}$.
We set that $\ket{0,p,a} $ is Grassmann even (or odd) if $\tilde{\gamma}^{11} = 1$  (or $\tilde{\gamma}^{11}=-1$).
We introduce the projection operators 
\begin{equation}
P_{ L} = \frac{1}{2} \left(1+\tilde{\gamma}^{11}  \right),
\qquad
P_{ R} = \frac{1}{2} \left(1-\tilde{\gamma}^{11}  \right)
\end{equation}
which respectively extract the $\tilde{\gamma}^{11}=\pm 1$ component from the ground states $\{\ket{0,p,a}\}$.
We often use the notation representing the ground state with $\tilde{\gamma}^{11} = \pm 1$ as
\begin{equation}
\ket{0,p,a:L} \equiv P_L \ket{0,p,a;\downarrow;\mbox{$-\frac{1}{2}$}},
\qquad
\ket{0,p,a:R} \equiv P_R \ket{0,p,a;\downarrow;\mbox{$-\frac{1}{2}$}}.
\label{eq:groundstLR}
\end{equation}
Note that $\ket{0,p,a:L} $ and $\ket{0,p,a:R} $ are Grassmann odd and even respectively.

On the spaces ${\cal H}_{\rm (NS)}$ and ${\cal H}_{\rm (R)}$, BRST operators $Q_{\rm NS}$ and 
$Q_{\rm R}$ are defined respectively.
They are divided by ghost zero modes as
\begin{eqnarray}
Q_{\rm NS} &=& \tilde{Q}+ c_0 L_0 + b_0 M  ,
\label{eq:QNS}
\\
Q_{\rm R}& =&\tilde{Q}+ c_0 L_0 + b_0 M  +\gamma_0\tilde{G}_0 +\beta_0 N -\gamma_0{}^2 b_0.
\label{eq:QR}
\end{eqnarray}
The definition of $\tilde{G}_0$, $\tilde{Q}$, $M$, and $N$ with other details of the matter and ghost operators are given in Appendix~A.


String fields $\Phi_{\rm NS}$ and $\hat{\Psi}_{\rm R}$ are expanded by the ghost number 1 string states within the spaces ${\cal H}_{\rm (NS)}$ and ${\cal H}_{\rm (R)}$. 
For each state, we assign a tensor or spinor-tensor field of appropriate Grassmann property
so that the string fields become Grassmann even in total,
that is, all fields within $\Phi_{\rm NS}$ are Grassmann even and those within $\hat{\Psi}_{\rm R}$ are Grassmann odd.
For future convenience, we separate those string fields into two parts and represent  
\begin{equation}
\Phi_{\rm NS} = \phi +c_0 \omega, \qquad
\hat{\Psi}_{\rm R} = \psi + (\gamma_0 + c_0\tilde{G}_0)  \chi
\label{eq:SF2}
\end{equation}
where $\phi, \omega \in \tilde{\cal F}^{\rm (NS)}$ and $\psi, \chi \in \tilde{\cal F}^{\rm (R)}$.
We also have to take into account the GSO projection~\cite{Gliozzi:1976qd}
which is defined by the G-parity operators 
\begin{equation}
G_{\rm NS} = (-1)^{F_{\rm NS}} ,\qquad 
G_{\rm R} = \tilde{\gamma}^{11} (-1)^{F_{\rm R}} 
\end{equation}
where
\begin{eqnarray}
{F_{\rm NS}}  &=& \sum_{r\ge \frac{1}{2}} \left( \psi_{-r}\cdot\psi_r +\beta_{-r}\gamma_r -\gamma_{-r}\beta_r  \right) -1,
\\
{F_{\rm R}}  &=& \sum_{n=1}^\infty \left( \psi_{-n}\cdot\psi_n +\beta_{-n}\gamma_n -\gamma_{-n}\beta_n \right)  +\gamma_0\beta_0.  
\end{eqnarray}
After performing the GSO projection and truncating the string fields to satisfy $G_{\rm NS, R}=1$,
the gauge invariant action for each sector is written as
\begin{equation}
S_{\rm NS}=\frac{1}{2}\left\langle  \Phi_{\rm NS} |Q_{\rm NS}   \Phi_{\rm NS} \right\rangle,
\qquad
S_{\rm R}=\frac{1}{2}\left \langle \overline{\hat{\Psi}}_{R}| c_0\delta'(\gamma_0) Q_{\rm R}  \hat{\Psi}_{R} \right\rangle
\end{equation}
where $\langle \Phi | = (| \Phi\rangle)^\dagger$ and $\langle \overline{\hat{\Psi}}_{\rm R} | = \langle {\hat{\Psi}_{\rm R} }| \tilde{\psi}_0{}^0$
and $\delta'(\gamma_0)=[\delta(\gamma_0), \beta_0]$.
Details of Hermitian conjugation and inner products are given in Appendix~A.
If we use the notation $|\cdot \rangle\!\rangle $ instead of $\ket{\cdot}$ as in eq.(\ref{eq:ketrel}), 
in terms of the inner product notation $\langle\!\langle \cdot || \cdot ||\cdot \rangle\!\rangle$ given by
eqs.(\ref{eq:innpro2NS})-(\ref{eq:innpro2}), 
the gauge invariant action for each sector is rewritten as
\begin{eqnarray}
S_{\rm NS} &=& -\frac{1}{2}\left(
\big\langle\!\big\langle \phi|| L_0 || \phi \big\rangle\!\big\rangle 
- \big\langle\!\big\langle \phi|| \tilde{Q} || \omega \big\rangle\!\big\rangle
-\big\langle\!\big\langle \omega||\tilde{Q}||\phi \big\rangle\!\big\rangle
-\big\langle\!\big\langle \omega||M ||\omega \big\rangle\!\big\rangle
\right),
\label{eq:S1NS}
\\
S_{\rm R} &=& \frac{1}{2}\left(
\big\langle\!\big\langle \bar{\psi}|| \tilde{G}_0 || \psi \big\rangle\!\big\rangle 
+ \big\langle\!\big\langle \bar{\psi}|| \tilde{Q}||\chi\big\rangle\!\big\rangle
+\big\langle\!\big\langle \bar{\chi}||\tilde{Q} || \psi \big\rangle\!\big\rangle
+ \frac{1}{2}\big\langle\!\big\langle \bar{\chi}|| (\tilde{G}_0 M+ M\tilde{G}_0) || \chi \big\rangle\!\big\rangle
\right).
\label{eq:S1R}
\end{eqnarray}
Here we have used the decomposition of $Q^{\rm NS, R}$:
\begin{eqnarray}
Q_{\rm NS} &=& \tilde{Q}+ c_0 L_0 + b_0 M  ,
\\
Q_{\rm R}& =&\tilde{Q}+ c_0 L_0 + b_0 M  +\gamma_0\tilde{G}_0 +\beta_0 N -\gamma_0{}^2 b_0.
\end{eqnarray}

Each action $S_{\rm NS}$ or $S_{\rm R}$ is invariant under the 
gauge transformation 
\begin{equation}
\delta \Phi_{\rm NS} = Q_{\rm NS}\Lambda ,
\qquad
\delta  \hat{\Psi}_{\rm R} = Q_{\rm R}\hat{\Lambda} .
\end{equation}
Here $\Lambda \in {\cal H}_{\rm NS}$ and $\hat{\Lambda} \in {\cal H}_{\rm R}$ 
are ghost number 0 string fields with $G_{\rm NS, R}=1$.  
Note that the gauge transformation is consistent with the GSO projection since
$[G_{\rm NS, R} ,Q_{\rm NS, R}]=0$ is satisfied.
By expressing the gauge parameter fields as  $\Lambda =\lambda + c_0 \rho$ and $\hat{\Lambda}=\lambda+(\gamma_0 +c_0 \tilde{G}_0) \xi$,
the gauge transformation is also rewritten as 
\begin{eqnarray}
\mbox{NS: }  && \delta \phi = \tilde{Q}\lambda +M \rho , \qquad \delta \omega = L_0 \lambda -\tilde{Q}\rho,
\label{eq:1gaugetrNS}
\\
\mbox{R: } && \delta \psi =\tilde{Q} \lambda + \frac{1}{2}(\tilde{G}_0 M+ M\tilde{G}_0) \xi,\quad
\delta{\chi} =\tilde{G}_0 \lambda +\tilde{Q}\xi.
\label{eq:1gaugetrR}
\end{eqnarray}

Note that the NS action is divided into two local gauge invariant parts $S_{\rm NS} = S_{{\rm NS}}^{\rm min} +S_{{\rm NS}}^{\rm auxiliary}$ consistently 
as in the case of bosonic theory~\cite{Asano:2006hk,Asano:2012qn}
whereas the action $S_{\rm R}$ does not have such decomposition. 
Note also that the sum $S_{\rm NS} + S_{\rm R}$ has ${\cal N}=1$ supersymmetry in $D=10$~\cite{Corrigan:1972tg,Corrigan:1973jz} 
and the explicit form of the supersymmetry transformation is given {e.g.,} in \cite{Kazama:1986cy}.
We will see the explicit form of the transformation for the massless part of the action in the next subsection.

\subsection{Massless action and the supersymmetry transformation}
We now explicitly see the properties of massless part of the action for later convenience. 
From now on, we fix $\alpha'=1$.
Massless fields are given by string states of level $N_{\rm level}=\frac{1}{2}$ for NS sector and $N_{\rm level}=0$ for R sector.
Thus the corresponding string fields in the form of eq.(\ref{eq:SF2}) after GSO projection are
\begin{eqnarray}
\phi_{m^2=0} &=& \int \!\frac{d^Dp}{ (2\pi)^D} \, \psi_{-\frac{1}{2}}^\mu\ket{0,p;\downarrow; -1} A_{\mu}(p)
,
\\
\omega_{m^2=0} &=& \int\! \frac{d^Dp}{ (2\pi)^D} \, {\frac{-1}{\sqrt{2}}}  \beta_{-\frac{1}{2}} \ket{0,p;\downarrow; -1} i C (p)
\end{eqnarray}
for NS sector and
\begin{equation}
\psi_{m^2=0} = \int \!\frac{d^Dp}{ (2\pi)^D} \, 
P_{L} \ket{0,p,a ;\downarrow; -\frac{1}{2}}  
\,\zeta_{a} (p) .
\end{equation}
for R sector.
Here the two fields $A_\mu$ and $C$ in the NS sector are Grassmann even real fields while $\zeta_{a}$ in the R sector is Grassmann odd left-handed real spinor field. 
The action for each sector can be given by substituting the above $\phi_{m^2=0}$ and 
$\omega_{m^2=0}$ into $S_{\rm NS}$, and $\psi_{m^2=0}$ (with $\chi_{m^2=0}=0$) into $S_{\rm R}$.
The result in the space-time representation is obtained after replacing $ip_{\mu} \rightarrow \partial_\mu $ as 
$$S_{\rm NS}^{m^2=0} = \int\! d^D\! x \, {\cal L}_{\rm NS}^{m^2=0} ,\qquad
S_{\rm R}^{m^2=0} = \int\! d^D\! x \, {\cal L}_{\rm R}^{m^2=0}$$
where 
\begin{eqnarray}
{\cal L}_{\rm NS}^{m^2=0} &=& \frac{1}{2} A_\mu \square A^\mu
-\frac{1}{2}C^2 + C \partial_\mu A^\mu
\nonumber\\
&=& -\frac{1}{4} F_{\mu\nu}F^{\mu\nu} -\frac{1}{2} (C-\partial_\mu A^\mu )^2,
\\
{\cal L}_{\rm R}^{m^2=0} &=& -\frac{1}{2} i \bar{\zeta} \partial\!\!\!/ \zeta
\end{eqnarray}
with $\bar{\zeta}=\zeta^\dagger\Gamma^0(= \zeta^{\rm T}\Gamma^0 )$.
The NS action $S_{\rm NS}^{m^2=0}$ is invariant under the gauge transformation
\begin{equation}
 \delta A_\mu = \partial_\mu \lambda , \qquad \delta C = \square
 \lambda
\end{equation}
which is read by substituting 
\begin{equation}
\lambda_{m^2=0} = \int\! \frac{d^Dp}{ (2\pi)^D} \,{\frac{1}{\sqrt{2}}} \beta_{-\frac{1}{2}} \ket{0,p;\downarrow; -1} i \lambda (p)
\end{equation}
into eq.(\ref{eq:1gaugetrNS}). 
Note that the NS part is divided into minimal physical part and the auxiliary field part  as
$ {\cal L}_{\rm NS}^{m^2=0} =   {\cal L}_{{\rm NS}}^{m^2=0, {\rm min}}+{\cal L}_{{\rm NS}}^{\rm{\rm auxiliary}}$
with 
\begin{equation}
{\cal L}_{{\rm NS}}^{m^2=0, {\rm min}}  =  -\frac{1}{4} F_{\mu\nu}F^{\mu\nu},\qquad
{\cal L}_{{\rm NS}}^{\rm{\rm auxiliary}} =-\frac{1}{2} (C-\partial_\mu A^\mu )^2  .
\end{equation}

The sum of the action $S_{\rm NS}^{m^2=0}+S_{\rm R}^{m^2=0}$ naturally represents the ${\cal N}=1$ supersymmetric gauge theory in $D=10$. 
The supersymmetric transformation for massless string fields is explicitly calculated by
using the relation given by eqs.(\ref{eq:susym01})-(\ref{eq:susym04}) as
\begin{eqnarray}
 \delta_\epsilon A_\mu = \bar{\epsilon} \Gamma_\mu \zeta ,\qquad \delta_{\epsilon} C = 0,
\qquad
\delta_\epsilon \zeta = i \partial\!\!\!/   A_\mu \Gamma^\mu\epsilon -i C \epsilon.
\label{eq:n1susytr}
\end{eqnarray}
where $\epsilon$ is a left-handed real spinor field.

\subsection{Gauge fixing: extension of the $a$-gauges}
We briefly comment on the gauge fixing problems for each sector by concentrating on the possible extension of $a$-gauge conditions, a class of covariant gauges,  introduced by the present authors for the bosonic string field theory in refs.\cite{Asano:2006hk, Asano:2012qn}. 

In the NS sector,
since the structure of the state space and the BRST operator with respect to ghost zero modes is similar to that of bosonic string theory, 
the $a$-gauge conditions~\cite{Asano:2006hk, Asano:2012qn} for the bosonic string field theory can be applied to the action $S_{\rm NS}$ straightforwardly.
That is, the gauge invariance of the action $S_{\rm NS}$ is fixed by the condition  
\begin{equation}
a \tilde{Q} \phi +M\omega =0
\label{eq:agaugecond}
\end{equation}
for an arbitrary real parameter $a\,(\ne 1)$.
For $a=0$ and $a=\infty$, the condition respectively corresponds to the Feynman-Siegel gauge and the Landau gauge.
The gauge-fixed action can be constructed by introducing the string fields with all the ghost numbers as in the case of bosonic string field theory.

In the R sector, on the other hand, direct extension of the $a$-gauges for bosonic theory 
to the action $S_{\rm R}$ cannot be performed since the contributions from the superghost zero modes $\gamma_0$ and $\beta_0$ should be taken into account.
For example, the condition such as $a \tilde{Q} \psi +M\tilde{G}_0\chi=0$ does not completely fix the gauge invariance of $S_{\rm R}$.
We can however show that the modified condition 
\begin{equation}
a (\tilde{Q} \psi - N\chi)  +M \tilde{G}_0 \chi =0
\label{eq:agaugecondR}
\end{equation}
for arbitrary $a\,(\ne 1)$ completely fixes the gauge invariance for $S_{\rm R}$.
We regard this class of gauge conditions as the (modified) $a$-gauge for the R-sector.
For $a=0$, the condition is reduced to $\chi=0$ and it coincides with the Feynman-Siegel gauge.
For $a=\infty$, unlike the bosonic or NS case, the condition does not have a particular property.

\section{Construction of extended superstring field theory}

\subsection{${\rm NS}^{ n}$ sector}
We construct the quadratic action of the extended string field theory for ${\rm NS}^{ n}$ sector.
This can be accomplished by applying the similar procedure used for constructing the  extended bosonic string field theory~\cite{Asano:2013rka}
since the state spaces for both theories have the same structure with respect to the ghost zero modes.
We first provide the direct product space of $n$ distinct open string state spaces for NS sector as 
\begin{equation}
{\cal H}_{{\rm NS}^{n}}(p) = {\cal H}_{\rm (NS)}^{(1)}(p) \otimes \cdots \otimes {\cal H}_{\rm (NS)}^{(n)}(p)
\end{equation}
and impose the condition on $\ket{f}_{{\rm NS}^{n}} \in {\cal H}_{{\rm NS}^{n}}(p) $ as
\begin{equation}
 (b_0^{(i)} -   b_0^{(j)} ) \ket{f}_{{\rm NS}^{n}} =0,\qquad
(L_0^{(i)} -   L_0^{(j)} ) \ket{f}_{{\rm NS}^{n}} =0
\label{eq:nNSbccond}
\end{equation}
for $i,j =1,\cdots,n$.
The latter equation of (\ref{eq:nNSbccond}) is equivalent to the level matching condition $N^{(i)}_{\rm level} -  N^{(j)}_{\rm level}=0 $ 
since the operator $L_0^{(i)}$ is written by using the level counting operator $N^{(i)}_{\rm level}$ as
$L_0^{(i)}=\ell^2 p^2/2 +N^{(i)}_{\rm level} -\frac{1}{2}$.
Here, $\ell$ is introduced as a scale factor instead of $\alpha'$ of the case of string theory $n=1$ or 2. In the following, we fix $\ell=\sqrt{2}$ and $\alpha_0^\mu=\sqrt{2}p^\mu$. 

We can construct the consistent extended string field theory on the restricted space ${\cal H}'_{{\rm NS}^{n}}\; (\subset {\cal H}_{{\rm NS}^{n}} )$ which is spanned by states satisfying the condition (\ref{eq:nNSbccond}).
This space can be divided into two parts 
\begin{equation}
{\cal H}'_{{\rm NS}^{n}} = \tilde{\cal F}_{{\rm NS}^{n}}  + \tilde{c}_0 \tilde{\cal F}_{{\rm NS}^{n}}
\label{eq:nNSspacediv}
\end{equation}
where $\tilde{c}_0 = \sum_{i=1}^n c_0^{(i)} $ 
and $\tilde{\cal F}_{{\rm NS}^{n}} (\subset  \tilde{\cal F}_{(\rm NS)}^{(1)} \otimes \cdots \otimes \tilde{\cal F}_{(\rm NS)}^{(n)} )$
does not include ghost zero modes.
Note that each $ \tilde{\cal F}_{(\rm NS)}^{(i)} (\subset{\cal H}_{\rm (NS)}^{(i)} ) $ is a copy of $ \tilde{\cal F}^{(\rm NS)}$ given in  (\ref{eq:Rstate}).
A state $ \ket{\tilde{f}}_{{\rm NS}^{n}} \in \tilde{\cal F}_{{\rm NS}^{n}}  $ can be represented by the form 
\begin{equation}
\ket{\tilde{f}}_{{\rm NS}^{n}} ={\cal O}_{\tilde{f}}  \ket{0,p;\downarrow^{1}\cdots\downarrow^{n}}_{{\rm NS}^{n}}
\end{equation}
where ${\cal O}_{\tilde{f}}  $ is an operator consisting of non-zero mode matter and ghost oscillators and 
\begin{equation}
 \ket{0,p;\downarrow^1\cdots\downarrow^n}_{{\rm NS}^{n}} =
\ket{0,p;\downarrow;-1 }_{\rm NS}^{(1)} \otimes \cdots\otimes \ket{0,p;\downarrow;-1 }_{\rm NS}^{(n)}
\;(= c_1^{(1)} \cdots c_1^{(n)} \ket{0,p}_{{\rm NS}^{n}} )
\end{equation}
is the ground state.
On the restricted space $ {\cal H}'_{{\rm NS}^{n}}$, the operators  $b_0^{(i)}$, $L_0^{(i)}$ and $N^{(i)}_{\rm level}$ do not depend on $i$
and we often omit the index $i$.
We also choose the GSO projection for every part $i$ as $G_{\rm NS}^{(i)}=+1$.
We denote by ${\cal H}^{'(+,\cdots,+)}_{{\rm NS}^{n}}$ such GSO projected space.

On the space $ {\cal H}'_{{\rm NS}^{n}}$, we define the extended BRST operator ${\cal Q}_{{\rm NS}^{n}}$ as the sum of $Q^{(i)}_{\rm{NS}}$ for all $i$'s :  
${\cal Q}_{{\rm NS}^{n}} = \sum_{i=1}^{n} Q^{(i)}_{\rm{NS}}$.
This operator can be divided by the ghost zero modes as
\begin{equation}
{\cal Q}_{{\rm NS}^{n}}  = \tilde{\cal Q} +\tilde{c}_0 L_0 +b_0 {\cal M}
\end{equation}
where 
\begin{equation}
\tilde{\cal Q} =\sum_{i=1}^{n} \tilde{Q}^{(i)}_{\rm{NS}} ,\qquad
{\cal M} = \sum_{i=1}^{n} M^{(i)}.
\end{equation}
For $D=10$, ${\cal Q}_{{\rm NS}^{n}}^2=0  $ holds.
As in the case of bosonic theory, 
${\cal M}$, ${\cal M}^- \,(=\sum_{i=1}^n M^{-(i)})$, and ${\cal M}_z\,(=\sum_{i=1}^n M_z^{(i)} ) $
constitute the SU(1,1) algebra (\ref{eq:su11}) for any $n$. 
Also, the relation 
\begin{equation}
 \tilde{\cal Q}^{2} + L_0 {\cal M}=0
\end{equation}
holds and the projection operators like ${\cal P}_k$, ${\cal W}_k$ and ${\cal S}_k$ defined for the bosonic extended string field theory~\cite{Asano:2013rka}  can also be defined for our ${\rm NS}^{ n}$ sector theory.
These operators play an important role in constructing the gauge fixed action for the $a$-gauges.

The inner product on the space ${\cal H}'_{{\rm NS}^{n}}$ is defined by 
\begin{equation}
 {}_{{\rm NS}^{n}}\langle  0,p' ; \downarrow^{1}\cdots\downarrow^{n}  | c_0^{(1)} \cdots c_0^{(n)} \ket{ 0,p;\downarrow^{1}\cdots\downarrow^{n}}_{{\rm NS}^{n}}
 = (-i)^{n(n-1)/2} (2 \pi)^D
\delta^{(D)}(p-p')
\label{eq:nNSinnpro}
\end{equation}
where
\begin{equation}
(\ket{0,p; \downarrow^{1}\cdots\downarrow^{n}   }_{{\rm NS}^{n}})^\dagger = {}_{{\rm NS}^{n}}\langle  0,p' ; \downarrow^{1}\cdots\downarrow^{n}  | 
= {}_{{\rm NS}^{n}}\langle 0,p'|
c_{-1}^{(n)} \cdots c_{-1}^{(1)} .
\end{equation}
The left-hand side of eq.(\ref{eq:nNSinnpro}) can be written as  
$
{}_{{\rm NS}^{n}}\langle 0,p'|
c_{-1}^{(n)} \cdots c_{-1}^{(1)} 
c_0^{(1)} \cdots c_0^{(n)} c_1^{(1)} \cdots c_1^{(n)} 
\ket{0,p}_{{\rm NS}^{n}}
$.
For convenience, we introduce the product ${}_{{\rm NS}^{n}}\langle\!\bra{0,p}|\cdot  |\ket{0,p}\!\rangle_{{\rm NS}^{n}}$ as 
\begin{equation}
i^{n(n-1)/2}\; {}_{{\rm NS}^{n}}\langle  0,p' ; \downarrow^{1}\cdots\downarrow^{n}  | {\cal O} c_0^{(1)} \cdots c_0^{(n)} \ket{ 0,p;\downarrow^{1}\cdots\downarrow^{n}}_{{\rm NS}^{n}}
\equiv  {}_{{\rm NS}^{n}}\langle\!\langle 0,p'|| {\cal O} | \ket{0,p}\!\rangle_{{\rm NS}^{n}} 
\label{eq:nNSinnpro2}
\end{equation}
with
\begin{equation}
 {}_{{\rm NS}^{n}}\langle\!\langle 0,p'|| 1 | \ket{0,p}\!\rangle_{{\rm NS}^{n}} = (2 \pi)^D \delta^{(D)}(p-p') 
\end{equation}
where ${\cal O}$ is an arbitrary operator consisting of matter and ghost oscillators without $b_0$ and $\tilde{c}_0$. 
Note that we can assume $(-1)^{|\cal O|}=1$ since the left-hand side of eq.(\ref{eq:nNSinnpro2}) vanishes otherwise.

The extended string field ${\Phi}_{{\rm NS}^{n}}$
for ${\rm NS}^{ n}$ sector is expanded by all states 
$\ket{f^k (p)} \in {\cal H}^{'(+,\cdots,+)}_{{\rm NS}^{n}}$ of ghost number $N_{\rm g}=n$
associated with the corresponding fields $\phi_{f^k}(p)$ as
\begin{equation} 
{\Phi}_{{\rm NS}^{n}} = \int \frac{d^{D}p}{(2\pi)^D} \sum_k  \ket{f^k (p)} \phi_{f^k}(p) 
\end{equation} 
where the Grassmann parity of ${\Phi}_{{\rm NS}^{n}}$ (and thus that for each field $\phi_{f^k}(p) $) should be even.
The quadratic action is given by
\begin{equation}
S_{{\rm NS}^{n}} = \frac{1}{2}(-1)^{n-1} i^{n(n-1)/2} 
\left\langle 
{\Phi}_{{\rm NS}^{n}} \big| c_0^{(1)} \cdots c_0^{(n-1)} \big| {\cal Q}_{{\rm NS}^{n}} {\Phi}_{{\rm NS}^{n}}  \right\rangle .
\end{equation}
If we divide the extended string field as 
\begin{equation}
 {\Phi}_{{\rm NS}^{n}} = \phi +\tilde{c}_0 \omega
\end{equation}
according to the division eq.(\ref{eq:nNSspacediv}) and use the notation given in eq.(\ref{eq:nNSinnpro2}) by 
replacing $\ket{\phi}  \rightarrow  \ket{\phi}\!\rangle$ and 
$\ket{\omega}\rightarrow \ket{\omega}\!\rangle$ 
as in the case of $n=1$ open string case, the action is represented by 
\begin{equation}
S_{{\rm NS}^{n}} = -\frac{1}{2}\left(
\big\langle\!\big\langle \phi|| L_0 || \phi \big\rangle\!\big\rangle 
- \big\langle\!\big\langle \phi||  \tilde{\cal Q}  || \omega \big\rangle\!\big\rangle 
-\big\langle\!\big\langle \omega|| \tilde{\cal Q} || \phi \big\rangle\!\big\rangle 
-\big\langle\!\big\langle \omega||{\cal M} || \omega \big\rangle\!\big\rangle 
\right)
\label{eq:nNSS2}
\end{equation}
which is exactly the same form as the action for open NS sector eq.(\ref{eq:S1NS}).
Note that we regard that the Grassmann parity of $|\phi\rangle\!\rangle$ and $|\omega\rangle\!\rangle$ is $(-1)^n$ and $(-1)^{n-1}$ respectively.

For $D=10$, this action is invariant under the gauge transformation
\begin{equation}
\delta {\Phi}_{{\rm NS}^{n}}  = {\cal Q}_{{\rm NS}^{n}} {\Lambda}_{{\rm NS}^{n}} 
\end{equation}
where $\Lambda_{{\rm NS}^{n}}$ is the ghost number $N_{\rm g} =n-1$ string field which is expanded by fields in  ${\cal H}^{'(+,\cdots,+)}_{{\rm NS}^{n}}$
with $N_{\rm g} =n - 1 $.
The transformation can be rewritten in terms of $\phi$ and $\chi$ as
\begin{equation}
\delta \phi = \tilde{\cal Q} \lambda +{\cal M} \rho , \qquad \delta \omega = L_0 \lambda - \tilde{\cal Q}  \rho,
\end{equation}
where $\Lambda_{{\rm NS}^{n}} = \lambda + \tilde{c}_0 \rho$.
We see that the structure of the action and the gauge invariance is parallel to that of bosonic string field theory in NS sector.
Thus, gauge-fixing procedures effective for open string field theory for NS sector 
can also be applied to our theory.

As in the case of bosonic theory, the action can be divided into two gauge invariant part:
\begin{equation}
S_{{\rm NS}^{n}} = S_{{\rm NS}^{n}}^{\rm min} + S_{{\rm NS}^{n}}^{\rm auxiliary} 
\label{eq:nNSSdiv2}
\end{equation}
where
\begin{eqnarray}
S_{{\rm NS}^{n}}^{\rm min} &=&
-\frac{1}{ 2}  \big\langle\!\big\langle \phi|| L_0 (1- {\cal P}_0 ) ||\phi \big\rangle\! \big\rangle,
\\
S_{{\rm NS}^{n}}^{\rm auxiliary} &=& 
\frac{1}{ 2}  \big\langle\!\big\langle \omega +{\cal W}_1 \tilde{\cal Q} \,\phi  ||{\cal M} ||\omega +{\cal W}_1 \tilde{\cal Q} \,\phi   \big\rangle\!\big\rangle.
\end{eqnarray}
Here ${\cal P}_0 = -\frac{1}{L_0}  \tilde{\cal Q} {\cal W}_1 \tilde{\cal Q} $
and ${\cal W}_1$, which plays a role of inverse of ${\cal M}$ on $\tilde{\cal F}_{{\rm NS}^{n}}^{\hat{N}_{\rm g}=1 } $, is defined in the same way as for the bosonic theory~\cite{Asano:2013rka}.
The auxiliary field part $S_{{\rm NS}^{n}}^{\rm auxiliary}$ can be decoupled from the physical action $S_{{\rm NS}^{n}}^{\rm min}$
if we only deal with the quadratic action. 
However, if we consider the supersymmetric action by combining with the fermionic action $S_{\mbox{\scriptsize {NS}$^{n\!-\!1}$-R}}$ given in the next subsection,
the auxiliary fields in  $S_{{\rm NS}^{n}}^{\rm auxiliary}$ are necessary for representing the supersymmetry transformation in a simple manner
as in the case of $n=1$ given in eq.(\ref{eq:n1susytr}). 
 
\subsection{${\rm NS}^{n-1}$-${\rm R}$ sector}
We now construct the extended string field theory for ${\rm NS}^{n-1}$-R sector.
We provide the  $n\!-\!1$ copies of the state space of NS sector and a state space of R sector and take the direct product as
\begin{equation}
{\cal H}_{\mbox{\scriptsize {NS}$^{n\!-\!1}$-R}} (p) =
 {\cal H}_{\rm (NS)}^{(1)} (p) \otimes \cdots \otimes {\cal H}_{\rm (NS)}^{(n-1)}(p)\otimes {\cal H}_{\rm (R)} (p).
\end{equation}
Then, we impose the condition on $ \ket{f}_{\mbox{\scriptsize {NS}$^{n\!-\!1}$-R}} \in {\cal H}_{\mbox{\scriptsize {NS}$^{n\!-\!1}$-R}} (p)$ as
\begin{equation}
 (b_0^{(i)} -   b_0^{(j)} )  \ket{f}_{\mbox{\scriptsize {NS}$^{n\!-\!1}$-R}} =0,\qquad
(L_0^{(i)} -   L_0^{(j)} )  \ket{f}_{\mbox{\scriptsize {NS}$^{n\!-\!1}$-R}}  =0
\label{eq:n-1NSRbccond}
\end{equation}
for $i,j =1,\cdots,n$ where we assign $n$ the space for R sector and consider ${\cal H}_{\rm (R)} = {\cal H}_{\rm (R)}^{(n)}  $. 
The latter equation restricts the level of each $i$ part as $ N^{(1)}_{\rm level} =\cdots=N^{(n-1)}_{\rm level} = N^{(n)}_{\rm level}+\frac{1}{2}  $.
We define the restricted space ${\cal H}'_{\mbox{\scriptsize {NS}$^{n\!-\!1}$-R}} $ by imposing the condition (\ref{eq:n-1NSRbccond})
on $ {\cal H}_{\mbox{\scriptsize {NS}$^{n\!-\!1}$-R}}$.
The space can be represented as 
\begin{equation}
 {\cal H}'_{\mbox{\scriptsize {NS}$^{n\!-\!1}$-R}} = \tilde{\cal F}_{\mbox{\scriptsize {NS}$^{n\!-\!1}$-R}}  +
 \left(\gamma_0 + \tilde{c}_0 \tilde{G}_0 \right) \tilde{\cal F}_{\mbox{\scriptsize {NS}$^{n\!-\!1}$-R}}
\end{equation}
where $\tilde{c}_0 = \sum_{i=1}^n c_0^{(i)} $ and 
$\tilde{\cal F}_{\mbox{\scriptsize {NS}$^{n\!-\!1}$-R}}  $ 
is a space of states without ghost and superghost zero modes in  ${\cal H}'_{\mbox{\scriptsize {NS}$^{n\!-\!1}$-R}}  $:
$
\tilde{\cal F}_{\mbox{\scriptsize {NS}$^{n\!-\!1}$-R}} =  {\cal H}'_{\mbox{\scriptsize {NS}$^{n\!-\!1}$-R}}  \cap
( \tilde{\cal F}_{\rm {(NS)}}^{(1)} \otimes\cdots \otimes  \tilde{\cal F}_{\rm {(NS)}}^{(n-1)}  \otimes \tilde{\cal F}_{\rm {(R)}}) .
$
As in the ${\rm NS}^{n}$ sector, note that the operation of $b_0^{(i)}$ and  $L_0^{(i)}$ on states in ${\cal H}'_{\mbox{\scriptsize {NS}$^{n\!-\!1}$-R}}  $ does not depend on $i$.
Also, we choose the GSO projection for all $n$ parts  as $G_{\rm{NS}}^{(i)}=G_{\rm{R}}=+1$ ($i=1,\cdots,n\!-\!1$)
and denote by ${\cal H}^{'(+,\cdots,+;+)}_{\mbox{\scriptsize {NS}$^{n\!-\!1}$-R}} (\subset  {\cal H}'_{\mbox{\scriptsize {NS}$^{n\!-\!1}$-R}})$ the GSO projected space.

On the space ${\cal H}'_{\mbox{\scriptsize {NS}$^{n\!-\!1}$-R}}  $, we define the extended BRST operator 
${\cal Q}_{\mbox{\scriptsize {NS}$^{n\!-\!1}$-R}} $ as 
\begin{equation}
{\cal Q}_{\mbox{\scriptsize {NS}$^{n\!-\!1}$-R}}  = \sum_{i=1}^{n-1} Q^{(i)}_{\rm{NS}} + Q_{\rm{R}} .
\end{equation}
This is decomposed by ghost zero modes as
\begin{equation}
{\cal Q}_{\mbox{\scriptsize {NS}$^{n\!-\!1}$-R}}    = \tilde{\cal Q} 
+ \tilde{c}_0 L_0 + b_0 {\cal M}  +\gamma_0\tilde{G}_0 +\beta_0 N -\gamma_0{}^2 b_0.
\end{equation}
Here $\tilde{\cal Q}$ and $\tilde{M}$ is the sum of all the corresponding operators in $n$ parts:
\begin{equation}
\tilde{\cal Q} =\sum_{i=1}^{n-1} \tilde{Q}^{(i)}_{\rm{NS}} + \tilde{Q}_{\rm{R}} 
\;\left( = \sum_{i=1}^{n} \tilde{Q}^{(i)}  \right) ,\qquad
{\cal M} = \sum_{i=1}^{n-1} M^{(i)}_{\rm NS} +   M_{\rm R}  
\;\left( = \sum_{i=1}^{n} M^{(i)}  \right).
\label{eq:QNSRdecomp}
\end{equation}
We see that the structure of $ {\cal Q}_{\mbox{\scriptsize {NS}$^{n\!-\!1}$-R}}  $ with respect to ghost zero modes is similar to that for open superstring 
theory in R sector.
In fact, the relation 
\begin{equation}
\tilde{\cal Q}^2+ L_0{\cal M} +\tilde{G}_0 N = 0 
\label{eq:squareQNSR}
\end{equation}
holds and the algebraic structure among the operators $\tilde{\cal Q}$ and $\tilde{M}$
along with ${\cal M}^-=\sum_{i=1}^n M^{(i)\,-}$ and ${\cal M}_z=\sum_{i=1}^n M_z^{(i)}$ is the same as that for the $n=1$ open superstring case.

The inner product on the space ${\cal H}'_{\mbox{\scriptsize {NS}$^{n\!-\!1}$-R}}  $ we use is given by
\begin{eqnarray}
 && {}_{(n\!-\!1,1)}\langle  0,p' ; \downarrow^{1}\cdots\downarrow^{n} , a':L(R) | c_0^{(1)} \cdots c_0^{(n)}  \delta(\gamma_0)  \tilde{\gamma}_{11}^n \ket{ 0,p;\downarrow^{1}\cdots\downarrow^{n},a:L(R)}_{(n\!-\!1,1)}
\nonumber\\
& & \hspace*{2cm} =  (-i)^{n(n-1)/2} (2 \pi)^D
\delta^{(D)}(p-p')\delta_{a,a'}
\label{eq:nNSRinnpro}
\end{eqnarray}
where 
\begin{equation}
\ket{ 0,p;\downarrow^{1}\cdots\downarrow^{n}:L(R)}_{(n-1,1)} = \ket{0,p; \downarrow^{1}\cdots\downarrow^{n-1}   }_{{\rm NS}^{n-1}}
\otimes \ket{0,p,a:L(R)}.
\end{equation}
Note that $\tilde{\gamma}_{11}^n \ket{0,p,a:L} = \ket{0,p,a:L}  $ and $\tilde{\gamma}_{11}^n \ket{0,p,a:R} =  (-1)^n \ket{0,p,a:R}  $.
We introduce another notation of inner product as before:
\begin{eqnarray}
&&\hspace*{-8mm} i^{n(n-1)/2}\; {}_{(n\!-\!1,1)}\langle  0,p' ; \downarrow^{1}\cdots\downarrow^{n}, a':L(R)  | {\cal O} c_0^{(1)} \cdots c_0^{(n)} 
\delta(\gamma_0)  \tilde{\gamma}_{11}^n \ket{ 0,p;\downarrow^{1}\cdots\downarrow^{n},a:L(R) }_{(n\!-\!1,1)}
\nonumber\\
& & \equiv \langle\!\langle 0,p',a'|| {\cal O} | \ket{0,p,a}\!\rangle 
\label{eq:nNSRinnpro2}
\end{eqnarray}
with 
\begin{equation}
\langle\!\langle 0,p',a' || 1  | \ket{0,p,a}\!\rangle = (2 \pi)^D \delta^{(D)}(p-p') \delta_{a,a'} .
\end{equation}
where ${\cal O}$ is a Grassmann even operator consisting only of non-zero mode oscillators as in the $n=1$ open superstring case.  
Note that the new notation of inner product is to be applied to the GSO projected space 
${\cal H}^{'(+,\cdots,+;+)}_{\mbox{\scriptsize {NS}$^{n\!-\!1}$-R}}$. 

The extended string field $\hat{\Psi}_{(n\!-\!1,1)}$ 
for the ${\rm NS}^{n-1}$-R sector has ghost number $N_{\rm g}=n$ and 
expanded by states 
$\ket{f^k (p)}$ 
 in  ${\cal H}^{'(+,\cdots,+;+)}_{\mbox{\scriptsize {NS}$^{n\!-\!1}$-R}}$ 
associated with Grassmann odd fields $\psi_{f^k}(p)$ as 
\begin{equation} 
\hat{\Psi}_{(n\!-\!1,1)} = \int \frac{d^{D}p}{(2\pi)^D} \sum_k  \ket{f^k (p)} \psi_{f^k}(p). 
\end{equation} 
Note that $ \hat{\Psi}_{(n\!-\!1,1)}$ is Grassmann even. 
The action is given by
\begin{equation}
S_{\mbox{\scriptsize {NS}$^{n\!-\!1}$-R}} = \frac{1}{2}
 (-1)^{n-1} i^{n(n-1)/2}
\left\langle 
\overline{\hat{\Psi}}_{(n\!-\!1,1)} \big| c_0^{(1)} \cdots c_0^{(n)} \delta'(\gamma_0)  {\cal Q}_{\mbox{\scriptsize {NS}$^{n\!-\!1}$-R}}   \hat{\Psi}_{(n\!-\!1,1)   }
\right\rangle .
\end{equation}
If we divide  $\hat{\Psi}_{(n\!-\!1,1)}$ by ghost zero modes as  
\begin{equation}
\hat{\Psi}_{(n\!-\!1,1)} = \psi + \left( \gamma_0+\tilde{c}_0 \tilde{G}_0\right)\chi,
\end{equation}
and use the notation given by the right-hand side of eq.(\ref{eq:nNSRinnpro}) after replacing 
$\ket{\psi} \rightarrow  \ket{\psi}\!\rangle$ and 
$\ket{\chi} \rightarrow \ket{\chi}\!\rangle$,
the action can be rewritten in a convenient form:
\begin{equation}
S_{\mbox{\scriptsize {NS}$^{n\!-\!1}$-R}} =  \frac{1}{2}\left(
\big\langle\!\big\langle \bar{\psi}|| \tilde{G}_0 || \psi \big\rangle\!\big\rangle 
+ \big\langle\!\big\langle \bar{\psi}|| \tilde{\cal Q}||\chi  \big\rangle\!\big\rangle
+\big\langle\!\big\langle \bar{\chi}||\tilde{\cal Q} || \psi  \big\rangle\!\big\rangle
+ \frac{1}{2}\big\langle\!\big\langle \bar{\chi}|| (\tilde{G}_0 {\cal M}+ {\cal M}\tilde{G}_0) || \chi  \big\rangle\!\big\rangle
\right) 
\label{eq:n-1NSRS2}
\end{equation}
where $\langle\!\langle\bar{\psi}|  = \langle\!\langle{\psi}|\tilde{\psi}_0^0$.  
This action is invariant under the gauge transformation
\begin{equation}
\delta \hat{\Psi}_{(n\!-\!1,1)} = {\cal Q}_{\mbox{\scriptsize {NS}$^{n\!-\!1}$-R}}  \hat{\Lambda}_{(n\!-\!1,1)}
\end{equation}
where $\hat{\Lambda}_{(n\!-\!1,1)}$ is ghost number $N_{\rm g}=n-1$.
This transformation is also expressed as 
\begin{equation}
\delta \psi=\tilde{\cal Q} \lambda + \frac{1}{2}(\tilde{G}_0 {\cal M}+ {\cal M}\tilde{G}_0) \xi,\quad
\delta{\chi} =\tilde{G}_0 \lambda +\tilde{\cal Q}\xi
\label{eq:nNSRgaugetr2}
\end{equation}
if we write $\hat{\Lambda}_{(n\!-\!1,1)}=\lambda+(\gamma_0+\tilde{c}_0 \tilde{G}_0) \xi $.
We see that the structure of the above gauge invariant action $S_{\mbox{\scriptsize {NS}$^{n\!-\!1}$-R}}$ is parallel to that of open superstring field theory in the R sector.
Thus we can straightforwardly apply the modified $a$-gauge condition eq.(\ref{eq:agaugecondR}) for the theory in R sector to our action 
$S_{\mbox{\scriptsize {NS}$^{n\!-\!1}$-R}} $ for general $n$.

\subsection{Massless part of  $S_{{\rm NS}^{n}}$ and $S_{\mbox{\scriptsize {NS}$^{n\!-\!1}$-R}}$}
We now see the structure of the massless part of the action for ${\rm NS}^{ n}$ or ${\rm NS}^{n-1}$-R sector 
given in the previous subsections.
The action for massless fields includes only a few kinds of oscillators:
$\psi^\mu_{-\frac{1}{2}}$, 
$\gamma_{-\frac{1}{2}}$ and $\beta_{-\frac{1}{2}}$ for NS part and 
$\psi_{0}^{\mu}$ for R part.
Thus, the operators $\tilde{\cal Q}$ and ${\cal M}$ on general massless extended string states are
reduced to 
\begin{equation}
 \tilde{\cal Q}_{{\rm NS}^{n}}^{m^2=0} = 
 \sum_{i=1}^{n} \sqrt{2}p_\mu 
 \left(\gamma_{-\frac{1}{2}}^{(i)} \psi_{\frac{1}{2}}^{\mu\,(i)}
 + \psi_{-\frac{1}{2}}^{\mu\,(i)}\gamma_{\frac{1}{2}}^{(i)} \right) ,
\qquad
{\cal M}_{{\rm NS}^{n}}^{m^2=0} = -2 \sum_{i=1}^{n} \gamma_{-\frac{1}{2}}^{(i)}\gamma_{\frac{1}{2}}^{(i)} 
\end{equation}
for ${\rm NS}^{ n}$ sector and 
\begin{equation}
 \tilde{\cal Q}_{(n\!-\!1,1)}^{m^2=0} = 
 \sum_{i=1}^{n-1} \sqrt{2}p_\mu 
 \left(\gamma_{-\frac{1}{2}}^{(i)} \psi_{\frac{1}{2}}^{\mu\,(i)}
 + \psi_{-\frac{1}{2}}^{\mu\,(i)}\gamma_{\frac{1}{2}}^{(i)} \right) ,
\qquad
{\cal M}_{(n\!-\!1,1)}^{m^2=0} = -2 \sum_{i=1}^{n-1} \gamma_{-\frac{1}{2}}^{(i)}\gamma_{\frac{1}{2}}^{(i)} 
\end{equation}
and 
\begin{equation}
N^{m^2=0}=0, \qquad \tilde{G}^{m^2=0}_0 =p_\mu \tilde{\psi}_0{}^\mu 
\end{equation}
for ${\rm NS}^{n-1}$-R sector. 
Note that $ ({\cal Q}_{{\rm NS}^{n}}^{m^2=0})^2= ({\cal Q}_{(n\!-\!1,1)}^{m^2=0})^2 =0 $ is satisfied not only for $D=10$ but also for arbitrary space-time dimension $D$.
Furthermore, for the R sector, since $N^{m^2=0}=0$, the relation
$$(\tilde{\cal Q}_{(n\!-\!1,1)}^{m^2=0})^2=-L_0  {\cal M}_{(n\!-\!1,1)}^{m^2=0}= - p^2{\cal M}_{(n\!-\!1,1)}^{m^2=0} $$ 
holds instead of general eq.(\ref{eq:squareQNSR}), 
and the modified $a$-gauge condition given by eq.(\ref{eq:agaugecondR}) is reduced to the form of the original $a$-gauge condition eq.(\ref{eq:agaugecond}) for bosonic or ${\rm NS}^{ n}$ sector theory.

\bigskip
We give some examples of gauge invariant actions for simple bosonic $n$-th rank tensor fields and fermionic $(n-1)$-th rank spinor-tensor fields 
extracted from $S_{{\rm NS}^{n}}^{m^2\!=\!0}$ and $S_{\mbox{\scriptsize {NS}$^{n\!-\!1}$-R}}^{m^2\!=\!0}$.

From the massless part of the action $S_{{\rm NS}^{n}}^{m^2\!=\!0}$ for the ${\rm NS}^{n}$ sector, 
as in the case of bosonic extended string field theory~\cite{Asano:2013rka}, we can extract consistent gauge invariant actions for $n$-th rank tensor fields of arbitrary symmetry classified by the Young diagrams of $n$ boxes.
For example, for totally symmetric field $A_{\mu_1\cdots \mu_n}=A_{(\mu_1\cdots \mu_n)}$, 
in order to obtain the corresponding action, we provide the extended string states of the form 
\begin{eqnarray}
 |\phi_{{\rm NS}^{n},\rm{sym}}^{m^2\!=\!0} \rangle\!\rangle &=& 
\int \!\! \frac{d^Dp}{(2\pi)^D}
\psi_{(1)}^{(\mu_1} \psi_{(2)}^{\mu_2}\cdots \psi_{(n)}^{\mu_n)}
\ket{0,p}\!\rangle_{{\rm NS}^{n}} \;
A_{ (\mu_1\mu_2\cdots \mu_n) } (p) 
\nonumber\\
& & \hspace*{-1.5cm}+
\sum_{i<{j}}(-1)^{i+j}
\psi_{(k_1)}^{(\mu_1} \psi_{(k_2)}^{\mu_2}\cdots\psi_{(k_{n-2})}^{\mu_{n-2}) }
(\gamma_{(i)}\beta_{(j)}- \beta_{(i)}\gamma_{(j)})
\ket{0,p}\!\rangle_{{\rm NS}^{n}} \;D_{ (\mu_1\cdots \mu_{n-2}) }(p) ,
\label{eq:Ansym}
\\
|\omega_{{\rm NS}^{n},\rm{sym}}^{m^2\!=\!0} \rangle\!\rangle &=& 
\int \!\! \frac{d^Dp}{(2\pi)^D}
\sum_{j=1}^n
 \frac{(-1)^j}{\sqrt{2}}
\psi_{(k_1)}^{(\mu_1} \psi_{(k_2)}^{\mu_2}\cdots\psi_{(k_{n-1})}^{\mu_{n-1}) }
\beta_{(j)}
\ket{0,p}\!\rangle_{{\rm NS}^{n}} \; 
i C_{ (\mu_1\mu_2\cdots \mu_{n-1}) }(p) 
\label{eq:Cnsym}
\end{eqnarray}
where we have omitted the subscripts $-\frac{1}{2}$ for oscillators 
and wrote {e.g.,} $\psi_{(i)}^\mu$ instead of $\psi^{(i)\,\mu}_{-\frac{1}{2}}$.  
Also, $n-2$ indices $k_r$'s are chosen so that $\{k_r, i,j\}=\{1,2,\cdots, n\}$ and $k_r<k_s$ for $r<s$ in eq.(\ref{eq:Ansym}), and $n-1$ $k_{r'}$'s in eq.(\ref{eq:Cnsym}) are chosen similarly:
$\{k_{r'}, j\}=\{1,2,\cdots, n\}$ as in ref.\cite{Asano:2013rka}. 
By substituting eqs.(\ref{eq:Ansym}) and (\ref{eq:Cnsym}) in eq.(\ref{eq:nNSS2}), we obtain
\begin{equation}
S_{{\rm NS}^{n}\,{\rm sym}}^{m^2\!=\!0} = \int d^D\!x\, 
{\cal L}_{n,\rm{sym}}
\label{eq:symnSb}
\end{equation}
in the $x$-representation after replacing $ip_\mu\rightarrow \partial_\mu$ with 
\begin{eqnarray}
{\cal L}_{n,\rm{sym}} &=&
\frac{1}{2}  A_{\mu_1\cdots \mu_n}  \square A^{\mu_1\cdots \mu_n} 
-\frac{1}{2}n(n-1)   D_{\mu_1\cdots \mu_{n-2}}  \square  D^{\mu_1\cdots \mu_{n-2}}  
-\frac{n}{2} C_{ \mu_1\cdots \mu_{n-1} } C^{ \mu_1\cdots \mu_{n-1} }
\nonumber\\
&& +n C_{ \mu_1\cdots \mu_{n-1}} \partial_\mu A^{\mu \mu_1\cdots \mu_{n-1}} 
- n(n-1) C_{\mu_1\cdots \mu_{n-1}} \partial^{(\mu_1} D^{\mu_2\cdots \mu_{n-1})}
\label{eq:nNSsym}
\end{eqnarray}
as expected.
This is divided into the minimal action part and the auxiliary field part:  
${\cal L}_{n,\rm{sym}}={\cal L}^{\rm min}_{n,\rm{sym}} + {\cal L}^{\rm auxiliary}_{n,\rm{sym.}}$ 
as 
\begin{eqnarray}
{\cal L}^{\rm min}_{n,\rm{sym}} &=&
\frac{1}{2}  A_{\mu_1\cdots \mu_n}  \square A^{\mu_1\cdots \mu_n} 
-n(n-1)   D_{\mu_1\cdots \mu_{n-2}}  \square  D^{\mu_1\cdots \mu_{n-2}}  
\nonumber\\
&&+\frac{n}{2} \partial_\mu A^{\mu \mu_1\cdots \mu_{n-1}} 
 \partial^\nu A_{\nu \mu_1\cdots \mu_{n-1}}  
+n(n-1) D_{\mu_1\cdots \mu_{n-2}} \partial_\mu \partial_\nu   
A^{\mu\nu\mu_1\cdots \mu_{n-2}}
\nonumber\\
&& + \frac{n(n-1)(n-2)}{2} \partial_\mu D^{\mu \mu_1\cdots \mu_{n-3}} 
 \partial^\nu D_{\nu \mu_1\cdots \mu_{n-3}}  ,
 \\
{\cal L}^{\rm auxiliary}_{n,\rm{sym.}} &=&
-\frac{n}{2} \left(
C_{ \mu_1\cdots \mu_{n-1}} -\partial^\mu A_{\mu\mu_1\cdots \mu_{n-1}}
+(n-1) \partial_{(\mu_1} D_{\mu_2\cdots \mu_{n-1})}
\right)^2.
\end{eqnarray}
These Lagrangians are respectively invariant under the gauge transformation
\begin{equation}
\delta A_{\mu_1\cdots \mu_n} =\partial_{(\mu_1} \lambda_{\mu_2\cdots \mu_n)},
\quad
\delta D_{\mu_1\cdots \mu_{n-1}} =\frac{1}{n} \partial^{\mu} \lambda_{\mu \mu_1\cdots \mu_{n-2}},
\quad
\delta C_{\mu_1\cdots \mu_{n-1}} = \square  \lambda_{\mu_1\cdots \mu_{n-1}}.
\end{equation}
Here, $\lambda_{\mu_1\cdots \mu_{n-1}} = \lambda_{(\mu_1\cdots \mu_{n-1})}$ which is given by the extended string field 
\begin{equation}
|\lambda_{{\rm NS}^{n},\rm{sym}}^{m^2\!=\!0} \rangle = 
\int \!\! \frac{d^Dp}{(2\pi)^D}
\sum_{j=1}^n
 \frac{(-1)^{j-1}}{\sqrt{2}}
\psi_{(k_1)}^{(\mu_1} \psi_{(k_2)}^{\mu_2}\cdots\psi_{(k_{n-1})}^{\mu_{n-1}) }
\beta_{(j)}
\ket{0,p}_{{\rm NS}^{n}} \; 
i \lambda_{ (\mu_1\mu_2\cdots \mu_{n-1}) }(p) .
\end{equation}
Similarly, gauge invariant action for other mixed symmetric $n$-th rank tensor fields can be 
extracted from $S_{{\rm NS}^{n}}$.
The result is parallel to the one obtained from the bosonic extended string field theory~\cite{Asano:2013rka} and we do not go into detail here.

On the other hand, from the action $S_{\mbox{\scriptsize {NS}$^{n\!-\!1}$-R}}^{m^2\!=\!0}$ for ${\rm NS}^{n-1}$-R sector, we can extract gauge invariant actions for various $(n\!-\!1)$-th rank left-handed real spinor-tensor fields.
The simplest one is for the totally symmetric spinor-tensor field $\phi^{(\mu_1\cdots \mu_{n\!-\!1})}$, for which the corresponding extended string field is given by
\begin{eqnarray}
 |\psi_{(n\!-\!1,1),\rm{sym}}^{m^2\!=\!0} \rangle\!\rangle &=& 
\int \!\! \frac{d^Dp}{(2\pi)^D}
\psi_{(1)}^{(\mu_1} \cdots \psi_{(n\!-\!1)}^{\mu_{n\!-\!1})}
\ket{0,p,a}\!\rangle\;
\phi^a_{(\mu_1\cdots \mu_{n\!-\!1})} (p) 
\nonumber\\
& & \hspace*{-1.5cm}+
\sum_{i<{j}}(-1)^{i+j}
\psi_{(k_1)}^{(\mu_1} \psi_{(k_2)}^{\mu_2}\cdots\psi_{(k_{n-3})}^{\mu_{n-3}) }
(\gamma_{(i)}\beta_{(j)}- \beta_{(i)}\gamma_{(j)})
\ket{0,p,a}\!\rangle \;\omega^a_{ (\mu_1\cdots \mu_{n-3}) }(p) ,
\label{eq:fnsym}
\\
|\chi_{(n\!-\!1,1), \rm{sym}}^{m^2\!=\!0} \rangle\!\rangle &=& 
\int \!\! \frac{d^Dp}{(2\pi)^D}
\sum_{j=1}^{n-1}
 \frac{(-1)^{j-n-1}}{\sqrt{2}(n\!-\!1)}
\psi_{(k_1)}^{(\mu_1} \cdots\psi_{(k_{n-2})}^{\mu_{n-2}) }
\beta_{(j)}
\ket{0,p,a}\!\rangle \; 
\chi^a_{ (\mu_1\cdots \mu_{n-2}) }(p) 
\label{eq:fchinsym}
\end{eqnarray}
where $\Gamma^{11} \phi=+\phi$, $\Gamma^{11} \omega=+\omega$, $\Gamma^{11} \chi=-\chi$ and the tensor indices for $\phi, \omega$, and $\chi$ fields are restricted to be totally symmetric. 
We have omitted the subscripts $-\frac{1}{2}$ for oscillators as before.  
Also, $n-3$ indices $k_r$'s are chosen so that $\{k_r, i,j\}=\{1,2,\cdots n\!-\!1\}$ and $k_r<k_s$ for $r<s$ in eq.(\ref{eq:fnsym}), and $n-2$ $k_{r'}$'s in eq.(\ref{eq:fchinsym}) are chosen similarly:
$\{k_{r'}, j\}=\{1,2,\cdots n\!-\!1\}$. 
By substituting eqs.(\ref{eq:fnsym}) and (\ref{eq:fchinsym}) into eq.(\ref{eq:n-1NSRS2}), we obtain 
\begin{eqnarray}
S_{\mbox{\scriptsize {NS}$^{n\!-\!1}$-R}}^{m^2\!=\!0} &=& \int d^D\!x \,
\frac{i}{2}\Big\{
- \bar\phi^{\mu_1\cdots \mu_{n\!-\!1}} /\!\!\!\partial\phi_{\mu_1\cdots \mu_{n\!-\!1}}
+(n-1)(n-2) \bar\omega^{\mu_1\cdots \mu_{n\!-\!3}} /\!\!\!\partial\omega_{\mu_1\cdots \mu_{n\!-\!3}}
\nonumber\\
&& \hspace*{-1cm}
-(n-2) \bar\omega^{\mu_1\cdots \mu_{n\!-\!3}} \partial_{\mu}\chi^{\mu}{}_{\mu_1\cdots \mu_{n\!-\!3}}
+ (n-2) \bar\chi^{\mu_1\cdots \mu_{n\!-\!2}} \partial_{(\mu_1}\omega_{\mu_2\cdots \mu_{n\!-\!2})}
\nonumber\\
&& \hspace*{-1cm}
+\bar\phi^{\mu_1\cdots \mu_{n\!-\!1}} \partial_{(\mu_1}\chi_{\mu_2\cdots \mu_{n\!-\!1})}
- \bar\chi_{\mu_1\cdots \mu_{n\!-\!2}} \partial_{\mu}\phi^{\mu\mu_1\cdots \mu_{n\!-\!2}}
+\frac{1}{n-1} \bar\chi^{\mu_1\cdots \mu_{n\!-\!2}} /\!\!\!\partial\chi_{\mu_1\cdots \mu_{n\!-\!2}}
\Big\}.
\label{eq:Sfsymn}
\end{eqnarray}
This action is equivalent to the so-called fermionic triplet action~\cite{Francia:2002pt, Sagnotti:2003qa} and
the physical degrees of freedom include those for fields with spin less than or equal to $n-\frac{1}{2}$. 
It is invariant under the gauge transformation
\begin{equation}
\delta \phi_{\mu_1\cdots \mu_{n\!-\!1}} =\partial_{(\mu_1} \lambda_{\mu_2\cdots \mu_{n\!-\!1})},
\quad
\delta \omega_{\mu_1\cdots \mu_{n\!-\!3}} =\frac{1}{n\!-\!1} \partial^{\mu} \lambda_{\mu \mu_1\cdots \mu_{n\!-\!3}},
\quad
\delta \chi_{\mu_1\cdots \mu_{n\!-\!2}} =  /\!\!\!\partial  \lambda_{\mu_1\cdots \mu_{n\!-\!2}}
\end{equation}
where $\Gamma^{11}\lambda = + \lambda$ and $\lambda_{\mu_1\cdots \mu_{n\!-\!2}} =\lambda_{(\mu_1\cdots \mu_{n\!-\!2})}$.
This gauge transformation is calculated by substituting 
\begin{equation}
|\lambda_{(n\!-\!1,1), \rm{sym}}^{m^2\!=\!0} \rangle\!\rangle =  
\int \!\! \frac{d^Dp}{(2\pi)^D}
\sum_{j=1}^{n-1}
 \frac{(-1)^{j-1}}{\sqrt{2}(n\!-\!1)}
\psi_{(k_1)}^{(\mu_1} \cdots\psi_{(k_{n-2})}^{\mu_{n-2}) }
\beta_{(j)}
\ket{0,p,a}\!\rangle \; i
\lambda^a_{ (\mu_1\cdots \mu_{n-2}) }(p) 
\end{equation}
into eq.(\ref{eq:nNSRgaugetr2}).

For the spinor field with totally anti-symmetric tensor indices $\phi_{[\mu_1\cdots \mu_{n\!-\!1}]} $,
the gauge invariant action is not simple enough compared to that for bosonic totally antisymmetric tensor field $B_{[\mu_1\cdots\mu_n]}$ whose minimal Lagrangian is given by the form 
${\cal L}\sim H_{\mu_1\cdots\mu_{n\!+\!1}}H^{\mu_1\cdots\mu_{n\!+\!1}} $ by the field strength $H$ of $B$.
In fact, gauge invariant action for $\phi_{[\mu_1\cdots \mu_{n\!-\!1}]}$ should contain the sequence of lower rank spinor-tensor fields $\phi'_{[\mu_1\cdots \mu_{n\!-\!3}]}$, $\phi''_{[\mu_1\cdots \mu_{n\!-\!5}]}$, $\cdots$ from the
$|\psi_{(n\!-\!1,1)}^{m^2\!=\!0}\rangle\!\rangle $ part, and 
$\chi_{[\mu_1\cdots \mu_{n\!-\!2}]}$, $\chi'_{[\mu_1\cdots \mu_{n\!-\!4}]}$, $\cdots$
from the $|\chi_{(n\!-\!1,1)}^{m^2\!=\!0}\rangle\!\rangle $ part.
Thus, the corresponding action includes $n-2$ kinds of lower rank fermionic fields other than the original $\phi_{[\mu_1\cdots \mu_{n\!-\!1}]}$ field in total.
This difference of properties between bosonic and fermionic actions results from the fact that the bosonic action can be divided into two gauge invariant parts as eq.(\ref{eq:nNSSdiv2}) while the fermionic action does not have such a consistent decomposition.
For example, the action for the 2nd rank spinor-tensor field $\phi_{[\mu\nu]} $ includes two more fields $\phi'$ and $\chi_\mu$.
It can be calculated from NS$^2$-R action $S_{\mbox{\scriptsize NS$^2$-R}}^{m^2\!=\!0}$ and the result is 
\begin{eqnarray}
S_{\mbox{\scriptsize NS$^2$-R,asym}}^{m^2\!=\!0} &=& \int d^Dx \,
\Big\{
- \frac{i}{2} \bar\phi^{[\mu\nu]} /\!\!\!\partial\phi_{[\mu\nu]}
- i \bar\phi' /\!\!\!\partial\phi'
+i  \bar\chi^{\mu} /\!\!\!\partial\chi_{\mu}
\nonumber\\
&& 
- i \bar\phi^{\mu\nu} \partial_{[\mu}\chi_{\nu]} -i \bar\phi'\partial_\mu\chi^{\mu}
+ i \bar{\chi}_{[\mu}\partial_{\nu]} \phi^{\nu\mu} + 2\bar{\chi}_\mu\partial^{\mu}\phi'
\Big\}.
\end{eqnarray}
Note that $\phi_{\mu\nu}$ and $\phi'$ are the left-handed ($\Gamma^{11}=1$) and $\chi$ is 
the right-handed ($\Gamma^{11}=-1$) real fermionic fields.
Similarly, in $S_{\mbox{\scriptsize ${\rm NS}^{n-1}$-R}}^{m^2\!=\!0}$, all the fields belonging to $|\psi^{m^2\!=\!0}\rangle\!\rangle$ and $|\chi^{m^2\!=\!0}\rangle\!\rangle$ are left-handed and right-handed respectively. 
The above action $S_{\mbox{\scriptsize NS$^2$-R,asym}}^{m^2\!=\!0}$ is invariant under the gauge transformation
\begin{equation}
\delta\phi_{\mu\nu} =2\partial_{[\mu}\lambda_{\nu]},
\quad
\delta\phi' =\partial_{\mu}\lambda^{\mu} -/\!\!\!\partial \xi,
\quad
\delta\chi_{\mu} =- /\!\!\!\partial \lambda_{\mu} +\partial_\mu\xi .
\end{equation}

For any other general $(n\!-\!1)$-th rank mixed-symmetric spinor-tensor field, we can similarly extract the consistent gauge invariant action from the total action $S_{\mbox{\scriptsize ${\rm NS}^{n-1}$-R}}^{m^2\!=\!0}$.
Note that such gauge invariant actions are classified by Young diagrams of $n\!-\!1$ boxes as in the case of tensor fields in the bosonic extended string field theory~\cite{Asano:2013rka}.

\subsection{Some comments}
We give some comments on the actions $S_{\mbox{\scriptsize ${\rm NS}^{ n}$}}$ and $S_{\mbox{\scriptsize ${\rm NS}^{n-1}$-R}}$ 
we have constructed above.
The basic structure of these actions does not depend on $n$ and the properties of these actions are parallel to those for $n=1$ open superstring actions in NS and R sectors.
For example, all the fields in  $S_{\mbox{\scriptsize ${\rm NS}^{ n}$}}$ and $S_{\mbox{\scriptsize ${\rm NS}^{n-1}$-R}}$ are respectively bosonic and fermionic fields, and the action $S_{\mbox{\scriptsize ${\rm NS}^{ n}$}}$ can be divided into two consistent gauge invariant parts, minimal action part and auxiliary fields part, while $S_{\mbox{\scriptsize ${\rm NS}^{n-1}$-R}} $ cannot.
Also, no-ghost theorem is trivial from the structure of the BRST operator in $D=10$
and the physical degrees of freedom can be easily obtained by counting the right-cone degrees of freedom.  

On the other hand, concrete physical spectrum depends on $n$.
For example, the number of physical degrees of freedom of massless spectrum for both ${\rm NS}^{ n}$ and ${\rm NS}^{n-1}$-R sectors is 
$8^n$ which is  respectively described by the $n$-th rank tensor field $A_{\hat{\mu}_1\cdots\hat{\mu}_n}$ and the $n$-th rank left-handed real spinor-tensor field $\phi_{\hat{\mu}_1\cdots\hat{\mu}_{n\!-\!1}}^{\hat{a}}$
where $\hat{\mu}_i$ represents the $D-2(=8)$ transverse degrees of freedom  and $\hat{a}$ the physical 8 degrees of freedom for the spinor index. 
Note that the equations of motion for ${\rm NS}^{n-1}$-R sector reduce the number of spinor indices by half and thus the number of massless degrees of freedom for both sectors coincides with each other.
There is also the infinite tower of massive fields for both sectors and the number of physical degrees of freedom also coincides with each other.
This is natural since the total action $S_n = S_{\mbox{\scriptsize ${\rm NS}^{ n}$}}+S_{\mbox{\scriptsize ${\rm NS}^{n-1}$-R}}$ should have ${\cal N}=1$  spacetime supersymmetry in $D=10$. 
For $n=1$, this is reduced to the usual supersymmetry for open superstring theory. 
For $n=2$, this symmetry between NS-NS and NS-R sectors is a part of ${\cal N}=2$ supersymmetry of closed superstring theory 
which is realized by including R-NS and R-R sectors in addition to the two sectors.
For general $n$, if we include all $2^n$ sectors consisting of totally $n$ NS and R parts in addition to ${\rm NS}^{ n}$ and ${\rm NS}^{n-1}$-R sectors, 
we might be able to construct the theory with ${\cal N}=n$ supersymmetry.

As to the construction of general sectors including more than two R parts, it is difficult to obtain an appropriate gauge invariant action in a straightforward way. 
For example, even for the R-R sector, it is known that we cannot construct the consistent gauge invariant action from the string field on
 the tensor product of two state spaces ${\cal H}_{\rm (R)} \otimes {\cal H}_{\rm (R)}$ with (or without) simple restriction such as $L_0-\bar{L}_0=0$ used for NS-NS or NS-R~\cite{Banks:1985xa, Ballestrero:1986aa}.
In general, it is possible to construct a consistent action by choosing a state space other than ${\cal H}_{\rm (R)} \otimes {\cal H}_{\rm (R)}$. 
One example is given in ref.\cite{Ballestrero:1986aa} where a certain gauge invariant action is constructed based on an asymmetrically chosen state space of the form ${\cal H}'_{\rm (R)} \otimes {\cal H}''_{\rm (R)}$.

We comment on the extension to general spacetime dimensions $D\ne 10$.
As we have seen above, since the BRST operator is nilpotent on general massless states for any $D$,
the massless part of the action is consistently applied for general $D$.
For fermionic massless part of the action $S^{m^2\!=\!0}_{\mbox{\scriptsize ${\rm NS}^{n-1}$-R}}$, however, 
we have to be careful about the type of spinors when we apply the action to a particular spacetime dimensions other than $D=10$.
For example, for $D=4$, the left-handed real spinor-tensor fields ($\phi$, $\omega$ and $\chi$)  appearing in the action eq.(\ref{eq:Sfsymn})
should be interpreted as general Weyl (or Majorana) spinors.
The similar reinterpretation is needed when we apply the actions to general $D$.

\section{Summary and discussion}
We have constructed the consistent quadratic gauge invariant actions for extended superstring field theory for ${\rm NS}^{ n}$ and ${\rm NS}^{n-1}$-R sectors.
The corresponding actions $S_{\mbox{\scriptsize ${\rm NS}^{ n}$}}$ and $S_{\mbox{\scriptsize ${\rm NS}^{n-1}$-R}}$ are extensions of those for NS and R sectors of open superstring field theory.
The massless spectrum of the theories for ${\rm NS}^{ n}$ and ${\rm NS}^{n-1}$-R sectors in general includes higher-spin gauge fields 
represented by general $n$-th rank tensor fields and spinor-tensor fields respectively.
From the actions, we are able to extract general quadratic gauge invariant actions for any type of tensor or spinor-tensor fields.

For massless fields, the actions for $n$-th rank tensor fields or spinor-tensor fields are classified by Young diagrams of $n$ boxes. 
The simplest examples are the actions eqs.(\ref{eq:symnSb}) and (\ref{eq:Sfsymn}) which are respectively for totally symmetric tensor and spinor-tensor fields. 
For a general mixed-symmetric tensor or spinor-tensor field represented by a certain Young diagram, 
we can obtain the explicit form of the action by extracting the corresponding part from $S_{\mbox{\scriptsize ${\rm NS}^{ n}$}}$ or $S_{\mbox{\scriptsize ${\rm NS}^{n-1}$-R}}$ and calculating the inner products of extended string states.
The actions for bosonic fields are the same as the ones obtained from the bosonic extended string field theory given in ref.\cite{Asano:2012qn} except for the space-time dimensions.
For fermionic fields, the structure of the actions are more complicated than that for bosonic fields as we have seen in the example of massless anti-symmetric field in the previous section.

Note that there have been various attempts of constructing consistent actions or field equations for general higher-spin fields in flat spacetime
({e.g.}, \cite{Francia:2002pt,  Sagnotti:2003qa, Buchbinder:2011xw, Buchbinder:2004gp, Campoleoni:2008jq, Campoleoni:2009gs, Reshetnyak:2012ec}).
From the perspective of constructing a consistent gauge invariant action for general higher-spin field represented by an unconstrained (spinor-)tensor field in flat spacetime, our method has certain advantages since the gauge invariance and the no-ghost theorem are trivial and covariant gauge fixing is straightforward.
Note also that the actions for massless fields can also be obtained from the tensionless limit of open superstring field theory for NS or R sector.

Since we have successfully constructed the free extended superstring field theories as well as the bosonic ones, 
our next task is to see whether we could also construct the consistent interaction terms.
For this purpose, we would like to analyze the structure of the known interaction terms for open and closed string field theory and study whether it is possible to extend them to the $n>2$ case.
This should be a challenging task since we do not even know what is the physical objects represented by the theories for $n>2$.
On the contrary, we may take more algebraic approach such as $A_{\infty}$/$L_{\infty}$.
We would like to tackle the problem for the simpler bosonic extended string field theory by using the gauge invariance structure as a hint.

\section*{Acknowledgements}
The work of M.K.~was supported in part by JSPS KAKENHI Grant Number 25287049.
\appendix 
\def\thesection{Appendix~\Alph{section}}
\renewcommand{\theequation}{\Alph{section}.\arabic{equation}}
\def\thesection{Appendix~\Alph{section}}
\setcounter{equation}{0}
\setcounter{figure}{0}
\def\thesection{Appendix~\Alph{section}}
\section{$\!\!$ Useful relations for open superstring states and operators}
\label{app0}
\setcounter{equation}{0}
\paragraph{Matter and ghost oscillators}
The (anti-)commutation relations of the matter and ghost oscillators for NS and R sectors are summarized as 
$[\alpha_{m}^\mu, \alpha_n^\nu]=m\eta^{\mu\nu} \delta_{m,-n}$, 
$\{b_m, c_n\}=\delta_{m,-n}$, 
\begin{equation}
 \mbox{NS:}\quad
\{\psi_r^\mu,  \psi_s^\nu \}=\eta^{\mu\nu} \delta_{r,-s},
\qquad 
 \mbox{R:}\quad
\{\psi_m^\mu,  \psi_n^\nu \}=\eta^{\mu\nu} \delta_{m,-n} 
\end{equation}
with $\eta_{\mu\nu}=\mbox{diag}(-1,+1,\cdots,+1)$ and 
\begin{equation}
 \mbox{NS:}\quad
[\gamma_r,\beta_s]=\delta_{r,-s},
\qquad 
 \mbox{R:}\quad
[\gamma_m,  \beta_n]= \delta_{m,-n} 
\end{equation}
where $m,n\in \mathbb{Z}$ and $r,s\in \mathbb{Z} +\frac{1}{2}$.
Hermitian properties are given by
\begin{eqnarray}
&&(\alpha_m^\mu)^\dagger=\alpha_{-m}^\mu,\quad b_m^\dagger =b_{-m},\quad
c_m^\dagger =c_{-m}, 
\\
&&
\beta_r^\dagger =-\beta_{-r},\quad \gamma_r^\dagger =\gamma_{-r}, 
 \quad
\beta_m^\dagger =-\beta_{-m},\quad \gamma_m^\dagger =\gamma_{-m}, 
\\
&&
(\psi_r^\mu)^\dagger=\psi_{-r}^\mu ,\qquad 
(\psi_m^\mu)^\dagger=\psi_{-m}^\mu\;\;(m\ne 0),\quad
\\
&&
(\psi_0^\mu)^\dagger=\eta_{\mu\nu} \psi_{0}^\nu 
\,(= 2 \psi_0^0 \psi_{0}^\mu \psi_0^0 ).
\end{eqnarray}
We often use  
$\tilde{\psi}_0^{\mu} (\equiv \sqrt{2} \psi_0^\mu ) $ instead of $\psi_0^\mu$ since the commutation relation for $\tilde{\psi}_0^{\mu}$ is the same as for Gamma matrices $ [\Gamma^{\mu},\Gamma^{\nu}]=2 \eta^{\mu\nu}  $.
\paragraph{Super Virasoro generators}
Matter part of super Virasoro generators is given by
\begin{equation}
L_n^{\rm (m)}  = \sum_{m=-\infty}^\infty \frac{1}{2}:\alpha_{-m}\cdot\alpha_{n+m} :
\,+
\!\! \sum_{ \substack{q+ \frac{1}{2}\in  \mathbb{Z}  ~{\rm (NS)} \\  q \in  \mathbb{Z}~{\rm (R)} }  }
\frac{1}{2}\left(\frac{1}{2}n+q \right) : \psi_{-q}\cdot \psi_{q+n}:
-\kappa_1 \delta_{n,0} 
\end{equation}
with $\kappa_1=0$ for NS and $\kappa_1=-\frac{5}{8}$ for R and
\begin{equation}
\mbox{NS:}\quad G_s^{\rm (m)} = \sum_{m\in \mathbb{Z}} \alpha_{-m}\cdot\psi_{s+m},
\qquad
\mbox{R:}\quad G_n^{\rm (m)}  = \sum_{m\in \mathbb{Z}} \alpha_{-m}\cdot\psi_{n+m}.
\end{equation}
Ghost part is 
\begin{equation}
L_n^{\rm (g)}  = \sum_{m=-\infty}^\infty (m-n) :b_{-m}c_{n+m} :
\,+
\!\! \sum_{ \substack{q+ \frac{1}{2}\in  \mathbb{Z}  ~{\rm (NS)} \\  q \in  \mathbb{Z}~{\rm (R)} }  }
\left(\frac{3}{2}n-q \right) : \gamma_{n-q} \beta_{q}: -\kappa_{2} \delta_{n,0} 
,
\end{equation}
with $\kappa_2=\frac{1}{2}$ for NS and $\kappa_2=\frac{5}{8}$ for R, and
\begin{eqnarray}
\mbox{NS:}&& G_s^{\rm (g)} = 
\sum_{m \in \mathbb{Z}} \left( \frac{1}{2}m-s \right) c_{-m} \beta_{m+s}
-2 \sum_{r+\frac{1}{2}\in \mathbb{Z}} \gamma_{-r}\beta_{s+r}
,
\\
\mbox{R:}&& G_n^{\rm (g)}  = 
\sum_{m \in \mathbb{Z}} \left( \frac{1}{2}m-n \right) c_{-m} \beta_{m+n}
-2 \sum_{m \in \mathbb{Z}} \gamma_{-m}\beta_{m+n}.
\end{eqnarray}
In total, $L_n= L_n^{\rm (m)} + L_n^{\rm (g)}$ and $G_q= G_q^{\rm (m)} + G_q^{\rm (g)}$ 
($q+\frac{1}{2} \in \mathbb{Z}$ for NS and  $q \in \mathbb{Z}$ for R)
and the total super Virasoro algebra in $D=10$ is given by
\begin{equation}
[L_{n},L_{m}] =(n-m) L_{n+m},\quad
[L_n,G_q] = \mbox{$\left( \frac{1}{2}n-q \right) $}G_{n+q},\quad
\{G_q,G_p \} = 2 L_{q+p}.
\end{equation}

\paragraph{BRST operators}
BRST operators $Q_{\rm NS}$ and $Q_{\rm R}$, which are nilpotent ($Q_{\rm NS}^2= Q_{\rm R}^2=0$) in 10-dimensional spacetime, 
are written as the form eqs.(\ref{eq:QNS}) and (\ref{eq:QR}).
The definitions of $M$ and  $\tilde{Q}$ in (\ref{eq:QNS}) and (\ref{eq:QR}) are given by
\begin{equation}
M = - \sum_{m=1}^\infty 2 m c_{-m} c_m
\;-\!\!\!
\sum_{\substack{q+ \frac{1}{2}\in  \mathbb{Z}_{>0}  ~{\rm (NS)} \\  q \in  \mathbb{Z}_{>0}~{\rm (R)} }}\!\!\! 2
\gamma_{-q}\gamma_q,
\end{equation}
\begin{eqnarray}
\tilde{Q} &=& \sum_{m\ne 0}  c_{-m} L_m^{\rm (m)}
-\!\!\frac{1}{2} 
\sum_{\substack{mn\ne 0 \\  m+n\ne 0 }} (m-n)\,:c_{-m} c_{-n} b_{n+m}: \;
+\!\!\! \sum_{\substack{q + \frac{1}{2} \in  \mathbb{Z}  ~{\rm (NS)} \\  q \in  \mathbb{Z}_{\ne 0}~{\rm (R)} }}
\gamma_{-q} G_{q}^{\rm (m)}
\nonumber\\
&& 
-  \!\!\!\sum_{ \substack{p+ q  \in  \mathbb{Z}_{\ne 0}  ~{\rm (NS)} \\  pq, p+q  \in  \mathbb{Z}_{\ne 0}~{\rm (R)} }}
\gamma_{-p}\gamma_{-q}b_{p+q}
+  \sum_{ \substack{m \in  \mathbb{Z}_{\ne 0}, q+\frac{1}{2}  \in  \mathbb{Z}  ~{\rm (NS)} \\  qm, q+m  \in  \mathbb{Z}_{\ne 0}~{\rm (R)} }}
\left( \frac{1}{2}m -q \right)
c_{-m}\gamma_{-q}\beta_{m+q}
\end{eqnarray}
where $ \mathbb{Z}_{>0}$ and $ \mathbb{Z}_{\ne 0}$  denote the sets of positive integers
and non-zero integers respectively.
The operators $N$ and $\tilde{G}_0$ are defined only for the R sector as
\begin{equation}
N=  \sum_{m \in \mathbb{Z} } \frac{3}{2}m c_{-m} \gamma_m,
\end{equation}
\begin{equation}
\tilde{G}_0 = \sum_{m \in \mathbb{Z} } \alpha_{-m}\cdot\psi_{m} 
- \sum_{m \in \mathbb{Z}_{\ne 0} }
\left(
\frac{m}{2} \beta_{-m}c_{m} +2 b_{m} \gamma_{-m}
\right).
\end{equation}
Note that $\tilde{G}_0$ is the non-zero mode part of $G_0 = G_{0}^{\rm (m)} +G_{0}^{\rm (g)}  $ and 
$\tilde{G}_0=G_0+2b_0c_0 $ holds.
The relation between $\tilde{G}_0$ and  $L_0$ is given by $\tilde{G}_0{}^2 (= G_0{}^2)=L_0 $.
Also, $L_0$ is written by the level counting operator $N_{\rm level}$ as
\begin{equation}
\mbox{NS:}\quad L_0 = \alpha' p^2 + N_{\rm level} -\frac{1}{2},\qquad
\mbox{R:}\quad L_0 = \alpha' p^2 + N_{\rm level} 
\end{equation}
where $\alpha_0^\mu =\sqrt{2 \alpha'}p^\mu$ and $N_{\rm level}$ is explicitly given by 
\begin{equation}
N_{\rm level}=\sum_{n\in \mathbb{Z}_{>0}} \left(\alpha_{-n}\cdot\alpha_n 
+n(c_{-n} b_n +b_{-n}c_n)
\right)\;
+\!\!\!\!\!\!\!
\sum_{\substack{q+ \frac{1}{2}\in  \mathbb{Z}_{>0}  ~{\rm (NS)} \\  q \in  \mathbb{Z}_{>0}~{\rm (R)} }}
\!\!\!\!\!
\left(
q \psi_{-q}\cdot \psi_q +q (-\gamma_{-q} \beta_q +\beta_{-q}\gamma_q)
\right).
\end{equation}
Commutation relations for these operators are 
\begin{equation}
[\tilde{Q},M] =[ M, N ] =   \{ \tilde{Q}, \tilde{G}_0\} =\{ \tilde{Q}, N\} = \{  \tilde{G}_0, N\} =  0,
\quad [M,\tilde{G}_0] =2 N.
\end{equation}
Also, $N^2=0$.
Nilpotent property of $Q$ is rewritten as 
\begin{equation}
\mbox{NS:}\quad \tilde{Q}^2 +ML_0 = 0,
\qquad
\mbox{R:}\quad \tilde{Q}^2 +ML_0 +\tilde{G}_0 N = 0 .
\end{equation}
Hermitian property of the above operators is given by
\begin{eqnarray}
\mbox{NS:}\quad  && \tilde{Q}^\dagger= \tilde{Q},\quad M^\dagger = M,
\\
\mbox{R:}\quad && \tilde{Q}^\dagger = \tilde{\psi}_{0}^0 \tilde{Q} \tilde{\psi}_{0}^0, \quad
M^\dagger = M,\quad
 \tilde{G}_0{}^\dagger = \tilde{\psi}_{0}^0 \tilde{G}_0\tilde{\psi}_{0}^0 , \quad N^\dagger=-N.
\end{eqnarray}

\paragraph{Ghost number operator and the SU(1,1) algebra}
Ghost number operator  $N_{\rm g}$ defined on  ${\cal H}_{\rm (NS)}$ and  ${\cal H}_{\rm (R)}$ is given by  
$N_{\rm g}= \hat{N}_{\rm g} +c_0b_0-\gamma_0\beta_0 +1$ with
\begin{equation}
\hat{N}_{\rm g} = 
\sum_{m\in \mathbb{Z}_{>0} } (c_{-m}b_m-b_{-m} c_m)
\;- \!\!\!\! \sum_{\substack{q+ \frac{1}{2}\in  \mathbb{Z}_{>0}  ~{\rm (NS)} \\  q \in  \mathbb{Z}_{>0}~{\rm (R)} }}
\!\!\!
(\gamma_{-q}\beta_q+\beta_{-q}\gamma_q ).
\end{equation}
Thus the ground states given by eqs.(\ref{eq:groundst}) have ghost number $N_{\rm g}=1$.
The operator $\hat{N}_{\rm g} $ counts the ghost number of the ghost non-zero mode part.
We can check the relations 
\begin{equation}
[\hat{N}_{\rm g},\tilde{Q}] =\tilde{Q} ,\quad
[\hat{N}_{\rm g},M] =2M, \quad
[\hat{N}_{\rm g},N] =2N.
\end{equation}

Note also that if we define 
\begin{equation}
M^- = - \sum_{m=1}^\infty \frac{1}{2m}  b_{-m} b_m
\;+\!\!\!
\sum_{\substack{q+ \frac{1}{2}\in  \mathbb{Z}_{>0}  ~{\rm (NS)} \\  q \in  \mathbb{Z}_{>0}~{\rm (R)} }}\!\!\! \frac{1}{2}
\beta_{-q}\beta_q
\end{equation}
and $M_z=\frac{1}{2}\hat{N}_{\rm g}$,
$M$, $M^-$ and $M_z$ constitute the SU(1,1) algebra 
\begin{equation}
[M,M^-]=2 M_z,\qquad [M_z,M]=M,\qquad [M_z,M^-]=-M^- 
\label{eq:su11}
\end{equation}
as in the case of bosonic string theory.

\paragraph{Inner products}
Inner product for spaces ${\cal H}_{\rm (NS)}$ and ${\cal H}_{\rm (R)}$ we use is given by the following rules:
\begin{equation}
\langle 0, p'\ket{0,p} = (2\pi)^D\delta^{(D)}(p-p'),\quad
\langle \downarrow |c_0 |\downarrow\rangle = \langle 0|c_{-1}c_0 c_1|0\rangle = 1,
\label{eq:innprog}
\end{equation}
\begin{equation}
\mbox{NS:}\quad \langle -1\ket{-1} =1,\qquad 
\mbox{R:}\quad 
 \langle \mbox{$-\frac{1}{2}$} |\delta (\gamma_0)\ket{-\frac{1}{2}} =1
.
\label{eq:innprog2}
 \end{equation}
Thus, 
\begin{equation}
{}_{\rm NS}\!\langle 0,p';\downarrow; -1 |c_0 \ket{0,p;\downarrow; -1}_{\rm NS} = 
(2\pi)^D\delta^{(D)}(p-p')
\label{eq:NSinnpro}
\end{equation}
for NS and
\begin{eqnarray}
\bra{0,p',a':L} c_0 \delta(\gamma_0) \ket{0,p,a:L} &=&(2\pi)^D\delta^{(D)}(p-p')\delta_{aa'}
\label{eq:RinnproL}
\\
\bra{0,p',a':R} c_0 \delta(\gamma_0) \ket{0,p,a:R} 
&=&
-(2\pi)^D\delta^{(D)}(p-p')\delta_{aa'}
\label{eq:RinnproR}
\end{eqnarray}
for R.
Here $\ket{0,p,a:L}=P_L\ket{0,p,a;\downarrow; -\frac{1}{2}} $ and $\ket{0,p,a:R}=P_R\ket{0,p,a;\downarrow; -\frac{1}{2}} $ 
as is defined in eq.(\ref{eq:groundstLR}).
Note that the minus sign of the right-hand side of (\ref{eq:RinnproR}) is due to the fact that $P_R \ket{0,p,a}$ is Grassmann odd.

For two states $\ket{f(p)}(={\cal O}_f \ket{0,p;\downarrow;-1}_{\rm NS})$ and  
$ \ket{g(p')}(={\cal O}_g \ket{0,p;\downarrow;-1}_{\rm NS})$ in the NS sector,
we define the inner product as 
\begin{equation}
\langle g(p')|f(p)\rangle =
{}_{\rm NS}\!\langle 0,p';\downarrow; -1 |
{\cal O}_g{}^\dagger {\cal O}_f
\ket{0,p;\downarrow; -1}_{\rm NS},
\end{equation}
which can be calculated by commutation relations for operators and the definition of inner product for ground state (\ref{eq:NSinnpro}).
We can define the inner product for any two states in the R sector in a similar manner.
Note that in the R sector, when constructing the quadratic action, 
we use $\langle \overline{f}|\equiv  \langle f| \tilde{\psi}_0{}^0$ 
instead of just taking the Hermitian conjugate as $ \langle f |$.
For a string field which is given by the infinite sum of string states with the corresponding fields, 
the relation between $ \ket{\Phi}$ and $\bra{\Phi}$ is 
\begin{equation}
\ket{\Phi} (= {\Phi})=\int\! \frac{d^Dp}{(2\pi)^D} \sum_i \ket{f_i(p)} \phi_{f_i}(p),
\qquad 
\bra{\Phi} =\int\!\frac{d^Dp}{(2\pi)^D} \sum_i \phi^*_{f_i}(-p)\bra{f_i(p)} .
\end{equation}

For convenience, we often use another notation of inner product which does not include $c_{0,\pm 1}$ nor $\gamma_0$ given by the right-hand side of the following equations:
\begin{eqnarray}
{}_{\rm NS}\!\langle 0,p';\downarrow; -1 |  {\cal O} c_0 \ket{0,p;\downarrow; -1}_{\rm NS} 
&\equiv& 
\langle\!\langle 0,p'||{\cal O} ||0,p\rangle\!\rangle ,
\label{eq:innpro2NS}
\\
\bra{0,p',a':L} {\cal O} c_0 \delta(\gamma_0) \ket{0,p,a:L}
&\equiv& 
\langle\!\langle 0,p'a'||{\cal O} ||0,p,a\rangle\!\rangle  ,
\label{eq:innpro2R1}
\\
\bra{0,p',a':R} {\cal O} c_0 \delta(\gamma_0) \ket{0,p,a:R}
&\equiv& 
- \langle\!\langle 0,p',a'||{\cal O} ||0,p,a\rangle\!\rangle 
\label{eq:innpro2R2}
\end{eqnarray}
where ${\cal O}$ is an arbitrary operator 
consisting only of matter and ghost oscillators without ghost zero modes and 
\begin{equation}
\langle\!\langle 0,p'|| 1 ||0,p\rangle\!\rangle = (2\pi)^D \delta^{(D)} (p-p') ,
\quad
\langle\!\langle 0,p',a'|| 1 ||0,p,a\rangle\!\rangle = (2\pi)^D \delta^{(D)} (p-p') \delta_{a,a'}
. 
\label{eq:innpro2}
\end{equation}
The right-hand sides of above three equations (\ref{eq:innpro2NS})-(\ref{eq:innpro2R2})
vanish when $(-1)^{|{\cal O}|}=-1$ if ${\cal O}$ consists only if matter and ghost oscillators.
For the R sector, eqs.(\ref{eq:innpro2R1}) and (\ref{eq:innpro2R2}) can be rewritten as a united form
\begin{equation}
\bra{0,p',a':L(R)} {\cal O} c_0 \delta(\gamma_0) \tilde{\gamma}^{11} \ket{0,p,a:L(R)}
\equiv 
\langle\!\langle 0,p'a'||{\cal O} ||0,p,a\rangle\!\rangle 
\end{equation}
and we use this new notation only for the GSO projected state space.
Note also that the essence of the above relations for both sectors is the relation 
$\bra{0,p'}c_{-1} {\cal O} c_0c_1 \ket{0,p}\equiv 
\langle\!\langle 0,p'||{\cal O} ||0,p \rangle\!\rangle $, i.e., 
the new notation of inner product corresponds to defining another state space spanned by 
states $ |f(p)\rangle\!\rangle$ instead of original $\ket{f(p)}$.
In concrete, for a state 
\begin{equation}
\ket{f(p)} = {\cal O}_f \ket{0,p;\downarrow;-1 }_{\rm NS} 
 \qquad
\mbox{or}
\qquad
\ket{f(p)} ={\cal O}_f \ket{0,p ,a:L(R) } ,
\end{equation}
$ |f(p)\rangle\!\rangle$ is defined as 
\begin{equation}
 | f(p) \rangle\!\rangle = (-1)^{|{\cal O}_f|} {\cal O}_f  |0,p \rangle\!\rangle 
\qquad
\mbox{or}
\qquad
 | f(p) \rangle\!\rangle = (-1)^{|{\cal O}_f|} {\cal O}_f  |0,p,a \rangle\!\rangle 
\label{eq:ketrel}
\end{equation}
where  $|0,p \rangle\!\rangle $ (or  $|0,p ,a\rangle\!\rangle $ ) is regarded as a new ground state 
satisfying $b_n|0,p \rangle\!\rangle = c_n|0,p \rangle\!\rangle =0 $ for $n>0$.
For example, the NS string field $\Phi_{\rm NS}=\phi + c_0 \omega$ can be expanded by 
\begin{eqnarray}
\phi \,(= \ket{\phi})& =& \int \!\! \frac{d^Dp}{(2\pi)^D} \sum_{i} {\cal O}_{f_i} \ket{0,p;\downarrow;-1}_{\rm NS} \phi_{f_i}(p),
\\
\omega \,(= \ket{\omega}) &=& \int \!\! \frac{d^Dp}{(2\pi)^D} \sum_{i} {\cal O}'_{g_i} \ket{0,p;\downarrow;-1}_{\rm NS} \omega_{g_i}(p)
\end{eqnarray}
where $(-1)^{|{\cal O}_{f_i}|} =-1$ and  $(-1)^{|{\cal O}'_{g_i}|} =1$.
Then, if we write 
\begin{eqnarray}
 |\phi \rangle\!\rangle & =& - \int \!\! \frac{d^Dp}{(2\pi)^D} \sum_{i} {\cal O}_{f_i} |0,p \rangle\!\rangle  \phi_{f_i}(p),
\\
  |\omega \rangle\!\rangle&=& \int \!\! \frac{d^Dp}{(2\pi)^D} \sum_{i} {\cal O}'_{g_i}  |0,p \rangle\!\rangle  \omega_{g_i}(p),
\end{eqnarray}
then the inner product can be expressed as 
\begin{eqnarray}
\bra{\phi}c_0\cdot\ket{\phi} &=& 
-\langle\!\langle \phi  || \cdot ||\phi \rangle\!\rangle  
\label{eq:innphi1}
\\
\bra{\omega }c_0 M \cdot\ket{\omega} &=& 
\langle\!\langle\omega | |\cdot ||\omega \rangle\!\rangle .
\label{eq:innphi2}
\end{eqnarray}
Note that we apply the new notation of inner product in the right-hand side of eqs.(\ref{eq:innphi1}) and (\ref{eq:innphi2}). 
The minus sign in the right-hand side of the former equation (\ref{eq:innphi1}) comes  from  the relation
\begin{equation}
\bra{0,p'} c_{-1} {\cal O}^\dagger_{f'} c_0 {\cal O}_f c_1 \ket{0,p} = 
- \bra{0,p'}  c_{-1} {\cal O}^\dagger_{f'} {\cal O}_f  c_0c_1 \ket{0,p}
=
- \langle\!\langle 0,p' || {\cal O}^\dagger_{f'} {\cal O}_f  || 0,p \rangle\!\rangle
\end{equation}
since $(-1)^{|{\cal O}_f|}=(-1)^{|{\cal O}_{f'}|}=-1$.

\paragraph{Supersymmetry transformation for massless fields}
Supersymmetry transformation can be constructed by the fermion emission vertex $W(z)$~\cite{Corrigan:1972tg, Corrigan:1973jz}.  
The explicit form of the transformation for the action $S_{\rm NS}+S_{\rm R}$ is given in ref.\cite{Kazama:1986cy} as
\begin{equation}
\left\{
\begin{array}{l}
\displaystyle
\delta_{\epsilon} \phi = \bar{W}_\epsilon \psi +  \bar{W}'_\epsilon \chi
\\
\displaystyle
\delta_{\epsilon} \omega = \bar{W}_{\epsilon} \tilde{G}_0 \chi
\end{array}
\right.,
\quad\quad 
\left\{
\begin{array}{l}
\displaystyle
\delta_{\epsilon} \psi = -\tilde{G}_0 {W}_\epsilon \phi +  {W}'_\epsilon \omega
\\
\displaystyle
\delta_{\epsilon} \chi = {W}_{\epsilon} \omega
\end{array}
\right.
\end{equation}
where ${W}_{\epsilon}$ and ${W}'_\epsilon$ are operators converting NS states into R states.
For massless ($N_{\rm level}=\frac{1}{2}$ for NS and $N_{\rm level}=0$ for R) string states, the transformation is given by
\begin{eqnarray}
{W}_\epsilon \psi_{-\frac{1}{2}} \ket{0,p;\downarrow; -1}_{\rm NS} &=&  \ket{0,p,a:R}  (\Gamma^\mu{\epsilon})_a,
\label{eq:susym01}
\\
{W}'_\epsilon \beta_{-\frac{1}{2}} \ket{0,p;\downarrow; -1}_{\rm NS} &=& \sqrt{2} \ket{0,p,a:L}  {\epsilon}_a,
\\
\bar{W}_\epsilon \ket{0,p,a:L} &=& \psi_{-\frac{1}{2}} \ket{0,p;\downarrow; -1}_{\rm NS} (\bar{\epsilon} \Gamma_\mu)_a,
\\
\bar{W}'_\epsilon \ket{0,p,a:R} &=& \sqrt{2} \gamma_{-\frac{1}{2}} \ket{0,p;\downarrow; -1}_{\rm NS} \bar{\epsilon}_a.
\label{eq:susym04}
\end{eqnarray}
Here $\epsilon$ is a Grassmann odd parameter real spinor field which satisfies $\Gamma^{11}\epsilon =\epsilon$ and 
$\epsilon^\ast =\epsilon$.



\end{document}